\documentclass[fleqn,usenatbib]{mnras}

\DeclareRobustCommand{\VAN}[3]{#2}
\let\VANthebibliography\thebibliography
\def\thebibliography{\DeclareRobustCommand{\VAN}[3]{##3}\VANthebibliography}

\usepackage{upgreek,slantsc}
\usepackage{ulem} 
\usepackage{cancel} 
\usepackage[dvipsnames]{xcolor}

\usepackage{makecell}

\usepackage{textcomp} 

\usepackage{graphicx}	
\usepackage{amsmath}	
\usepackage{amssymb}	

\definecolor{perhaps_teal}{rgb}{0, 0.5, 0.5}
\definecolor{bluish}{rgb}{0.4, 0.4, 1}
\definecolor{mypurple}{rgb}{0.7,0.3,0.8}

\newcommand{\luke}[1]{\textcolor{cyan}{#1}}

\newcommand\an{Astron.\ Nachr.}

\newcommand\gafd{Geophys.\ Astrophys.\ Fluid Dyn.}

\newcommand{\const}{\text{const}  {}} 
\newcommand{\dd}{\mathrm{d}} 
\newcommand{\e}{\mathrm{e}} 
\newcommand\uvec[1]{{\widehat{\vec{#1}}}}        
\newcommand{\cra}{_\text{cr}} 
\newcommand{\el}{_\text{e}} 
\newcommand{\disc}{_\text{d}} 

\newcommand{\mean}[1]{\overline{#1}}
\newcommand{\mB}{\mean{B}} 
\newcommand{\tB}[1]{B} 
\newcommand{\sound}{_\text{s}} 
\newcommand{\rms}{_\text{rms}} 

\newcommand{\HI}{\ion{H}{I}}   
\newcommand{\eq}{_\mathrm{eq}}  
\newcommand{\SFR}{\text{SFR}} 

\renewcommand{\vec}[1]{\boldsymbol{#1}}
\newcommand{\cm}{\,\text{cm}}    
\newcommand{\km}{\,\text{km}}    
\newcommand{\p}{\,\text{pc}}     
\newcommand{\kpc}{\,\text{kpc}}  

\newcommand{\msun}{\,\text{M}_\odot}  

\newcommand{\GHz}{\,\text{GHz}}  
\newcommand{\Hz}{\,\text{Hz}}    
\newcommand{\s}{\,\text{s}}      
\newcommand{\yr}{\,\text{yr}}    
\newcommand{\Myr}{\,\text{Myr}}  

\newcommand{\kms}{\,\km\s^{-1}}    

\newcommand{\G}{\,\text{G}}      
\newcommand{\mkG}{\,\text{{\textmu}G}} 

\newcommand{\erg}{\,\text{erg}}  
\newcommand{\GeV}{\,\text{GeV}}  
\newcommand{\K}{\,\text{K}}      
\newcommand{\mJyb}{\,\text{mJy/beam}}    
\newcommand{\W}{\,\text{W}}      

\newcommand{\gf}{\textsc{galform}} 
\newcommand{\magnet}{\textsc{magnetizer}} 
\title[The RLF of star-forming galaxies and its cosmological evolution]{Understanding the radio luminosity function of star-forming galaxies and its cosmological evolution}

\author[C.~Jose et al.]{Charles Jose$^1$\thanks{charles.jose@cusat.ac.in, lchamandy@niser.ac.in, anvar.shukurov@ncl.ac.uk, kandu@iucaa.in, luizfelippeSR@gmail.com, c.m.baugh@durham.ac.uk}, Luke Chamandy$^{2,3}$, Anvar Shukurov$^4$,
Kandaswamy Subramanian$^{5,6}$,
\newauthor 
Luiz Felippe S.~Rodrigues$^{7}$ and Carlton M.~Baugh$^8$\\
$^1$Department of Physics, CUSAT, Cochin, 682022, India\\
$^2$National Institute of Science Education and Research, An OCC of Homi Bhabha National Institute, Bhubaneswar 752050, Odisha, India\\
$^3$Department of Physics and Astronomy, University of Rochester, Rochester NY 14627, USA\\
$^4$School of Mathematics, Statistics and Physics, Newcastle University, Newcastle upon Tyne, NE1 7RU, UK\\
$^5$IUCAA, Post Bag 4, Ganeshkhind, Pune 411007, India\\
$^{6}$Department of Physics, Ashoka University, Rajiv Gandhi Education City, Rai, Sonipat 131029, Haryana, India\\
$^7$HAL24K Agri, Uitmeentsestraat 19, 6987 CX Giesbeek, Netherlands\\
$^8$Institute for Computational Cosmology, Department of Physics, University of Durham, South Road, Durham DH1 3LE, UK}
\date{Accepted XXX. Received YYY; in original form ZZZ}
\pubyear{2024}
\begin{document}
\pagerange{\pageref{firstpage}--\pageref{lastpage}}
\maketitle

\begin{abstract}
We explore the redshift evolution of the radio luminosity function (RLF) of star-forming galaxies using \gf, a semi-analytic model of galaxy formation and a dynamo model of the magnetic field evolving in a galaxy. Assuming energy equipartition between the magnetic field and cosmic rays, we derive the synchrotron luminosity of each sample galaxy. In a model where the turbulent speed is correlated with the star formation rate, the RLF is in fair agreement with observations in the redshift range $0 \leq z \leq 2$. At larger redshifts, the structure of galaxies, their interstellar matter and turbulence appear to be rather different from those at $z\lesssim2$, so that the turbulence and magnetic field models applicable at low redshifts become inadequate. The strong redshift evolution of the RLF at $0 \leq z \leq 2$ can be attributed to an increased number, at high redshift, of galaxies with large disc volumes and strong magnetic fields. On the other hand, in models where the turbulent speed is a constant or an explicit function of $z$, the observed redshift evolution of the RLF is poorly captured. The evolution of the interstellar turbulence and outflow parameters appear to be major (but not the only) drivers of the RLF changes. We find that both the small- and large-scale magnetic fields contribute to the RLF but the small-scale field dominates at high redshifts. Polarisation observations will therefore be important to distinguish these two components and understand better the evolution of galaxies and their nonthermal constituents.
\end{abstract}


\defcitealias{Rodrigues+19}{R19}

\section{Introduction}
\label{firstpage}

The galaxy luminosity function (LF) -- the comoving number density of galaxies as a function of their magnitude at a given wavelength --  is one of the primary observables used to probe the physics of formation and evolution of galaxies. The observed LFs and their redshift evolution have been measured, to various degrees of detail and reliability, for a wide range of wavelengths from the UV to the radio. When assessed in the framework of a theoretical model of galaxy formation, the LFs provide key insights into the evolution of star formation and feedback processes, the nature of and conditions in the ISM (consisting of gas, dust, magnetic fields, and cosmic rays), and dynamical processes such as galaxy mergers. Two major approaches used to model galaxy formation are: (i)~semi-analytic models (SAMGFs) that account for the complex baryonic processes in evolving dark matter (DM) haloes derived from $N$-body simulations using a combination of analytic approximations and numerical prescriptions, (ii)~hydrodynamic or magnetohydrodynamic (MHD) simulations based on fundamental dynamical equations, that include subgrid models to account for unresolved physical processes(often similar to those used in SAMGFs).
There are several galaxy formation models \citep{trayford+2015, ana_maarten+2022, croton+2006, samui+2007, Jose+2013,Lacey+16} that interpret the ultra-violet (UV) \citep{wyder_UVLF+2005, page_UVLF+2021}, optical \citep{blanton_opticalLF+2003, loveday_opticalLF+2022} and infrared (IR) \citep{dunne_IRLF+2011, Gruppioni+2013, marchetti_IRLF+2016} LFs and shed light on the evolution of obscured star formation rate (SFR), stellar mass ($M_\ast$), ISM, dust and feedback processes in galaxy populations. 

Complementary to the UV, optical and IR LFs, is the radio LF (RLF) of star-forming galaxies (SFGs) at the emission frequency $1.4\GHz$ obtained with the Very Large Array (VLA) and the Westerbork Synthesis Radio Telescope (WSRT) \citep{adams_WSRT+2019, condon_VLA+1998}. Recent estimates of RLFs of SFGs at $1.4\GHz$ include those of \citet{Sadler+02, condon+2002, best+2005, Mauch_Sadler_2007, padovani+2011} and \citet{condon+2019} for the local Universe and from \citet{smolcic+2009, pracy+2016_LF,Novak_2017, bonato+2021, malefahlo+2021, enina+2022} and \citet{vanderVlugt_2022} for redshifts up to $z \simeq 4.5$. Moreover, several ongoing and upcoming deep surveys using present and next-generation radio telescopes, including the Square Kilometre Array (SKA), will improve our knowledge of the RLFs of SFGs to higher redshifts and fainter luminosities with unprecedented accuracy \citep{norris+2013, adams_WSRT+2019}. For example, deep surveys using SKA Phase-I will detect sub-$\upmu\text{Jy}$ SFGs up to $z \simeq 6$ which will help to measure RLFs down to luminosities more than an order of magnitude lower than what is available now \citep{jarvis+2015}.

Synchrotron emission dominates the radio continuum at $1.4\GHz$ \citep{ginzburg+1965}, so models of the galactic magnetic field and cosmic ray propagation should be the starting points in a prediction of the RLF. MHD simulations of evolving galaxies \citep[e.g.,][]{Liu_mhd+2022, pfrommer+2022} cannot provide the statistically significant galaxy samples needed to compute RLFs. Moreover, limited spatial resolution prevents such simulations from probing scales as small as 1--$100\p$, which is crucial for capturing dynamo processes at the turbulent and global scales, and hence realistic magnetic fields and, consequently, any realistic distribution of cosmic rays controlled by them \citep[see section~13.14 of][for a review]{ss21}. Turbulent magnetic fields, produced by the fluctuation dynamo, supernova shock fronts, and tangling of the large-scale magnetic field, require a spatial resolution of order $1\p$ to be realistically reproduced \citep{GMLKS21,GMLKLS23}, whereas the mean-field dynamo, that generates large-scale galactic magnetic fields, relies, apart from overall rotation, on the density stratification at scales of order $100\p$ \citep[chapter~11 of ][]{ss21}. MHD simulations at an adequate resolution are only available for local ISM regions of order a few kiloparsecs in size. \citep[e.g.][]{Gressel+08a,GSFSM13a,GSFSM13b,HSSFG17,Gent+23b}, and subgrid dynamo models are not yet available despite an effort to relate the statistical properties of magnetic fields in nearby galaxies to theories \citep[\citealt{VEBSF15,CST16}; see][for a review]{Beck+19}.

Therefore, at present, combining SAMGFs with a model for the evolution of magnetic fields and cosmic rays is arguably the only viable theoretical approach for modelling RLFs. 

The earliest effort towards incorporating galactic dynamo theory into a hierarchical model for galaxy formation in cold DM cosmology to produce galactic magnetic fields in a cosmological volume-sized sample of galaxies was by \citet{Rodrigues+15}. \citet{Rodrigues+19} (hereafter \citetalias{Rodrigues+19}) extended this model into a computational framework called  \magnet\ \citep{Rodrigues+Chamandy20} which simulates magnetic fields in individual galaxies as they evolve. In particular, these authors coupled the \magnet\  with the output of the \gf~SAMGF of \citet{Lacey+16} and \citet{gonzalez+2014} to obtain simulated magnetic fields for a large sample of galaxies in the redshift range $0\leq z\leq6$ and compared the predictions of mean-field strengths and pitch angles with observations of local galaxies, finding reasonable agreement.

The present work extends the model of \citetalias{Rodrigues+19}, to derive the RLF of SFGs at the rest-frame frequency of $1.4\GHz$ over cosmic history. We calculate the total radio luminosity due to the synchrotron emission based on our magnetic field and cosmic ray models. Any significant differences of this work from the model of \citetalias{Rodrigues+19} are mentioned in the text.

Key quantities that affect the amplification and sustenance of galactic magnetic fields are the gas density and root-mean-square (rms) turbulent speed, as well as the galactic rotation and stratification (the gas scale height). Some observational studies indicate that the gas velocity dispersion in galaxies (which includes a contribution of the turbulence) is correlated with the SFR surface density (SFRD) \citep{green+2010, lehnert+2013, moiseev+2015} while other authors find a correlation with the global SFR \citep{Genzel+11,Varidel+16,Varidel_Croom+20,Zhou+17,Ubler_Genzel+19}.

We investigate how the predicted shape and redshift evolution of the RLF are affected by various phenomenological ISM turbulence models and compare the results with observations. By contrast, past works have probed the radio luminosity of isolated template galaxies and used the radio-FIR correlation to deduce the RLF \citep[e.g.][]{werhahn+2021, pfrommer+2022,vollmer+2022, Schober+23}.

The RLFs obtained from observations contain contributions from both active galactic nuclei (AGN) and SFGs \citep{Mauch_Sadler_2007, vanderVlugt_2022}. The physical nature and processes responsible for the radio emission in these objects are rather different and should be considered separately. These contributions can be separated, with varying degrees of confidence, using multi-wavelength observations to estimate the RLF of each. In this work, we model the RLF of star-forming disc galaxies only. 

This paper is organized as follows. In Section~\ref{model}, we discuss our models to compute the RLF of galaxies at $1.4\GHz$. We present the predictions for the magnetic field in Section~\ref{MFiEG}, and for the RLF in Section~\ref{sec_radioLF} along with a comparison with observational data and a discussion of the roles of various physical processes involved. Our results are put into a wider context in Section~\ref{Disc} and conclusions are formulated in Section~\ref{sec_conclusion}. Throughout this work, we use cosmological parameters based on seven-year WMAP data \citep{wmap7}, which are also used in the \gf~model of \cite{Lacey+16}. We use the same parameter values in \magnet\ as \citetalias{Rodrigues+19} unless otherwise stated. 

\section{The Model}\label{model}
The RLFs presented in this work are the products of a three-level model. In the first level, the output of the \gf\ SAMGF of \citet{Lacey+16} is used to derive the properties of a representative galaxy population of about $7 \times 10^5$ disc galaxies in a comoving volume of $6 \times 10^6$ Mpc$^3$ at each redshift in the redshift range $0 \leq \ z \leq 6$ (corresponding to 3.5\% of the total volume of the N-body simulation used to provide halo merger histories), with stellar masses between about $ 10^8\msun$ and $10^{12}\msun$. In the second level, \magnet\ is used to couple the galactic dynamo theory and the galaxy formation model to predict the evolution of both turbulent and global magnetic fields as a function of time and galactocentric radius in each galaxy. To avoid excessive and unnecessary complications, all galaxy properties are assumed to be axisymmetric; the dominance of axially symmetric large-scale magnetic fields in nearby galaxies is well established \citep[e.g.,][]{Beck+19}. In the final third stage, the synchrotron emission at $1.4\GHz$ is computed for each galaxy to derive the RLF over a wide redshift range.  

\subsection{Galaxy formation model}
The \gf\ version that we use \citep[] [see also \citealt{Cole2000, Baugh+2005, Bower+2006}]{Lacey+16} incorporates theoretically and empirically motivated models for complex physical processes (star formation, supernova, and AGN feedback, various dynamical processes, the evolution of the gas, stars, and dust, etc.) that baryons undergo in DM haloes. The assembly histories of haloes are described by halo merger trees extracted from $N$-body simulations \mbox{\citep{Guo+13}.}
The model successfully reproduces the observed K-band, optical, near-infrared, and far-UV luminosity functions of SFGs, far-IR number counts and \HI\ and stellar mass functions over a wide range of redshifts along with the Tully--Fisher, metallicity--luminosity and size--luminosity relations at $z = 0$. 
Nearly all radio-bright SFGs are spirals \mbox{\citep{Sadler+02, condon+2002}}. Therefore, as in \mbox{\citetalias{Rodrigues+19}}, we only consider disc galaxies in the output of \gf, which are defined as galaxies where the bulge mass is less than half the total stellar mass.

\gf\ outputs several global properties of galaxies at every snapshot of the $N$-body simulation, like the SFR, stellar mass $M_\ast$, the cold gas masses of the disc and the bulge, and the disc half-mass radius ($r_{1/2}$), which is assumed to be the same for stars and gas. Several ISM parameters for \magnet\ are derived from these \gf\ outputs in the same way as in \citetalias{Rodrigues+19}, as briefly discussed below, but we refer the reader to that paper for further details.

\citetalias{Rodrigues+19} focused on the galactic mean fields and therefore selected galaxies with a large disc that can host the mean-field dynamo in a relatively large volume. However, as we discuss below, random magnetic fields generated independently of the mean-field dynamo action make a significant contribution to the total-intensity RLF, and central parts of galaxies can be very bright in the synchrotron. Therefore, we have changed the sample selection criterion and include all galaxies that have a gas disc, either large or small. The motivation is that a higher local gas density in the disc would lead to star formation, turbulence, and, hence, significant magnetic and cosmic ray energy densities.

\subsubsection{Derived galactic quantities}\label{DGQ}
To obtain the RLF, we need to deduce the distribution of the gas density and the magnetic field along the galactocentric distance $r$ and distance $Z$ from the mid-plane in each galaxy, as well as its rotation curve and the gas scale height.

The galaxy rotation curve, $V(r)$, is computed by assuming that galaxies have a thin stellar disc with an exponential surface mass density profile, a bulge with the \cite{Hernquist_1990} profile and a DM halo with an adiabatically contracted Navarro--Frenk--White density profile \citep{NFW:1997}. The rotation curve is then used to determine the galaxy angular velocity, $\Omega(r)$, and the rotational shear rate, $S(r) =  r\,\dd\Omega/\dd r$.

As in \citetalias{Rodrigues+19}, we adjust the half-mass radius $r_{1/2}$ of the \gf\ galaxies to reproduce the observed stellar mass--$r_{1/2}$ relation of \cite{Lange+16} and $r\disc=2.7 r_{1/2}$ is adopted as the maximum gas disc radius for computing quantities like the volume-averaged magnetic field strength.

The surface mass densities of the stars and gas in the disc are both assumed to have an exponential radial profile with the scale length $r_\mathrm{s}= r_{1/2}/a$, where  $a\approx1.68$.\footnote{To obtain $a$, we note that $M(r)=2\pi\Sigma_0 r_\mathrm{s}^2\int_0^{r/r_\mathrm{s}}xe^{-x}dx$ for the mass of either stars or gas within the cylindrical radius $r$, where $\Sigma_0$ is the surface density at $r=0$. As noted in the text, the scale length $r_\mathrm{s}$ is assumed to be the same for both stars and total gas.Since $\int_0^\infty xe^{-x}\,\dd x=1$, we have $M_\infty\equiv\lim_{r\rightarrow\infty}M(r)=2\pi \Sigma_0 r_\mathrm{s}^2$.Now, $M(r_{1/2})=M_\infty/2$ by definition, which leads to $\int_0^a x\e^{-x}\,\dd x=1/2$ and $a$ follows.} As in  \citetalias{Rodrigues+19}, the disc gas mass is separated into diffuse and `molecular' phases. The ratio of the surface densities of the diffuse and molecular gas components is computed using the empirical relation from  \citet{Blitz+Rosolowsky04, Blitz+Rosolowsky06}. The density distributions of diffuse gas, molecular gas and stars perpendicular to the galactic disc are assumed to be exponential with different scale heights. Thus, for example, for the diffuse gas density we have
\begin{equation}
  \rho_\mathrm{d}(r,Z) = \rho_\mathrm{d0}(r) \e^{- |Z|/h_\mathrm{d}(r)}\,,
  \label{eq_dennsity_z}
\end{equation}
where $Z$ is the distance from the galaxy mid-plane, $\rho_\mathrm{d0}(r)$ is the mid-plane density and $h\disc(r)$ is the density scale height, both functions of the galactocentric radius $r$, determined assuming that the ISM is in hydrostatic equilibrium with a total pressure that includes the thermal, turbulent, magnetic and cosmic-ray contributions. We use empirical relations for the scale heights of the molecular gas ($0.032r_\mathrm{s}$) and stars ($0.1r_\mathrm{s}$) \citep{Kregel+2002, Licquia_Newman_2016}. 

\subsubsection{Interstellar turbulence}
\label{ism_turbulence}
Galactic magnetic fields are amplified and sustained by a turbulent dynamo. Thus, magnetic field models depend on certain statistical properties of the turbulence, like the correlation length $l$, correlation time $\tau$, and rms turbulent velocity $v$. As in \citetalias{Rodrigues+19}, we adopt $l(r) = \min[100\p, h_\mathrm{d}(r)]$ and $\tau=l/v$ \citep[for estimates of interstellar turbulence parameters, see][]{Chamandy+Shukurov20}.

\begin{figure}
\centering
\includegraphics[width=0.45\textwidth]{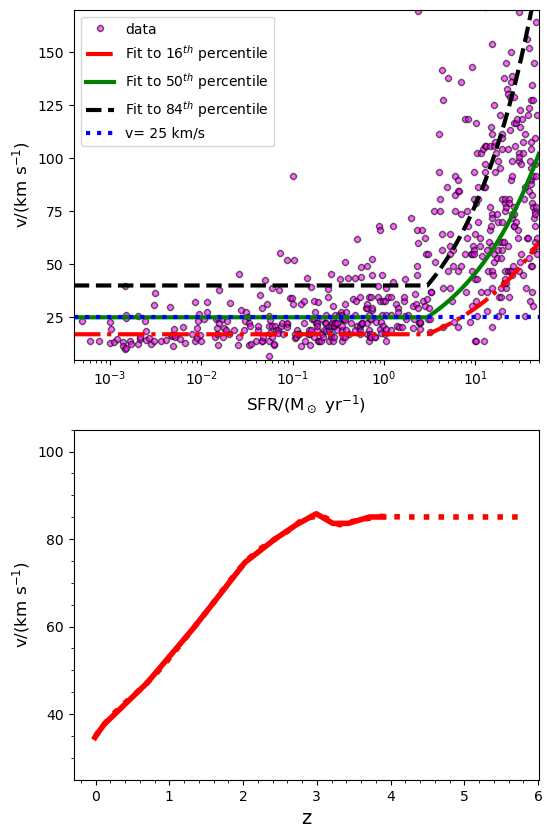}
\caption[ Velocity dispersion]{
Top panel: the observed three-dimensional velocity dispersion as a function of SFR (\citealt{Krumholz_18} and references therein, circles) and the best-fitting curves of the form of equation \eqref{eq_v0} to the median (solid/green) and the 16th (dash-dot/red) and 84th (dashed black) percentiles. The dotted blue line corresponds to $v=25\kms$. Bottom panel: The turbulent speed of the cold interstellar gas ($T\leq10\K$) as a function of redshift from the \textit{EAGLE} cosmological simulations for $z \leq 4$ \citep[solid line:][]{Jimenez_Lagos+23}, extrapolated to $z>4$ as a constant (dotted). 
}
\label{fig_turbv0}
\end{figure}

Thus far, the \magnet\ model described is the same as that of \citetalias{Rodrigues+19}. However, \citetalias{Rodrigues+19} assumed that the rms turbulent speed is $v=10\kms$, close to the sound speed in the warm gas and independent of the redshift and SFR. However, the gas velocity dispersion (which includes an uncertain contribution from the turbulence) appears to depend on the intensity of star formation (see Fig.~\ref{fig_turbv0} and the text below). Given this uncertainty, we considered three alternative prescriptions for determining the turbulent speed. 

Interstellar turbulence has various drivers, including star formation and the associated supernova activity, gravitational instability  \citep{Krumholz_16, Krumholz_18}, cosmological gas accretion, and galaxy mergers \citep{Jimenez_Lagos+23, Ginzburg_Dekel+22}. Some models \citep[e.g.,][]{Krumholz_18,Ginzburg_Dekel+22} predict that the  turbulent speed is independent of the $\SFR$ when $\SFR<\SFR_0$ with a certain threshold $\SFR_0$ and  increases with the $\SFR$ otherwise. The velocity dispersion data, consistent with more recent observations \citep{Varidel_Croom+20, Yu_Bian+21}, are shown in the upper panel of Fig.~\ref{fig_turbv0}, and we use the form 
\begin{equation}
  v = v_0
  \begin{cases}
    1\,, & \text{if } \SFR \leq \SFR_0\,, \\
    (\SFR/\SFR_0)^c\,, & \text{otherwise}, 
  \end{cases}\label{eq_v0}
\end{equation}
with certain constants $v_0$ and  $c$. As suggested by fig.~3 of \citet{Yu_Bian+21}, we adopt $\SFR_0=3\msun\yr^{-1}$ and estimate $v_0=25\kms$ and $c=0.50$ by fitting the median of the relation shown in Fig.~\ref{fig_turbv0}. The fitted dependence is shown as a solid curve in Fig.~\ref{fig_turbv0}, where the one-dimensional velocity dispersion $\sigma$ has been converted to the three-dimensional speed $v=\sigma\sqrt3$ assuming the turbulence to be isotropic. Further, $v$ is assumed to be independent of $r$ for simplicity.

The data points in Fig.~\ref{fig_turbv0} have a large scatter. To account for this, we fit equation~\eqref{eq_v0} to the $16^\text{th} $ and $84^\text{th}$ percentiles of $v$ as a function of SFR, to obtain $v_0 = 17\kms$ and $ c=0.45$ for the $16^\text{th}$ percentile and $v_0 = 40\kms$ and $c=0.55$ for the $84^\text{th}$ percentile, also shown in Fig.~\ref{fig_turbv0}. These measures of the scatter are used to estimate the degree of uncertainty of the predicted luminosity functions. We note that the SFR-$v$ relation is assumed to be independent of redshift in this model, which we adopt as the fiducial model.

The turbulent speed of $v_0=25\kms$ can be understood as the volume average of the rms turbulent speed in the multi-phase ISM. As an illustration, if the turbulence is transonic in both warm and hot phases, and their respective fractional volumes and sound speeds are $f_\text{w}=0.9$, $c_\text{w}=10\kms$ and $f_\text{h}=0.1$, $c_\text{h}=100\kms$, the volume average follows as $(f_\text{w}c_\text{w}^2+f_\text{h}c_\text{h}^2)^{1/2}\approx33\kms$. We note, however, that the fractional volumes of the ISM phases are likely to vary between galaxies with different SFRs and between various locations within a given galaxy.

As an alternative, we consider a model with a constant turbulent speed $v=25\kms$ suggested at lower SFRs by Fig.~\ref{fig_turbv0}, independently of the SFR and redshift (whereas \citetalias{Rodrigues+19} used $v=10\kms$). For the gas number density $0.1\cm^{-3}$ of the warm interstellar gas, the magnetic field strength at energy equipartition with the turbulence is then about $4\mkG$, close to the strength of the random magnetic field observed in nearby spiral galaxies (see Section~\ref{sec:random}). The increase in the velocity dispersion at large SFR in Fig.~\ref{fig_turbv0} can be interpreted to arise partly from chaotic galactic fountains and winds, which are especially vigorous in galaxies with high SFR. This model is called $v$25-R19 below.
 
The velocity dispersion is observed to be correlated with other parameters in addition to the SFR. These include the gas fraction, DM halo mass and the stellar mass \citep[e.g.,][]{Krumholz_18, Ginzburg_Dekel+22}. This implies that the SFR--$v$ correlation evolves with the redshift, which has been verified by the observations of \citet{Ubler_Genzel+19} and also found by \cite{Jimenez_Lagos+23} using the \textit{EAGLE} hydrodynamic simulations \citep{Schaye_Eagle_2015}. Therefore, we also consider a third model, referred to as $v(z)$-J23, where $v$ is an explicit function of $z$, as given by the REF-L100 model of \cite{Jimenez_Lagos+23} and plotted in the bottom panel of Fig.~\ref{fig_turbv0}. In this model, $v$ does not depend on the SFR explicitly. 
\begin{figure}
\centering
\includegraphics[width=0.45\textwidth]{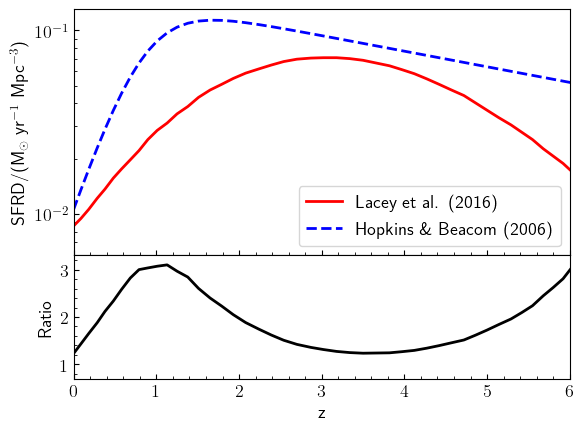}
\caption{Top panel: the best-fitting dependence of the observed  SFRD on the redshift from \citet{Hopkins_Beacom_06} (dashed/blue) and the SFRD from  \citet{Lacey+16} (solid/red).  
Bottom panel: the ratio of the SFRDs shown in the top panel.}
\label{figSFRoffset}
\end{figure}
\subsubsection{Correction to the SFR from used in \gf} \label{secSFRoffset}
In the version of \gf ~described in \cite{Lacey+16}, stars form out of cold gas in the disc and spheroid, and the corresponding SFR is proportional to the mass of the cold, molecular gas. The ratio of molecular to atomic gas surface densities is assumed to depend on the pressure in the mid-plane of the disk, using the empirical prescription of \citet{Blitz+Rosolowsky06} (see section 3.4 of \citet{Lacey+16} for details). The comoving SFR density (SFRD), the SFR per unit comoving cosmological volume (not to be confused with the star formation density within an individual galaxy), from \cite{Lacey+16} is shown as a function of redshift in the top panel of Fig.~\ref{figSFRoffset}. Also shown is the best fit to the observed redshift evolution of the SFRD,  derived by \citet{Hopkins_Beacom_06} from multi-wavelength photometric and spectroscopic observations across a wide range of wavelengths from X-ray to radio. Both the theoretical and observed SFRDs are computed after assuming the IMF of \citet{Kennicutt_83} \citep[see section~6.3 of][for details]{Lacey+16}. It is clear from Fig.~\ref{figSFRoffset} that, while observations and predictions agree quite well at $z=0$, the predicted SFRD lies below the fit to the observations at all redshifts, with their ratio shown in the lower panel. To account for this discrepancy and reconcile the SFR with that observed, we multiply the SFR of galaxies from \gf\ at a given redshift with this ratio, taken at the same redshift.  
\subsection{Magnetic field} \label{sec:Magnetizer} 
Large-scale (mean) $\vec{\mB}$ and small-scale (random) $\vec{b}$ magnetic fields contribute comparably to the galactic synchrotron emission (here and elsewhere, a bar above a variable denotes ensemble or volume average). In nearby galaxies, the ratio of their mean energy densities, $\mean{b^2}/\mB^2$, often significantly exceeds unity \citep{Beck16,Beck+19}, but this is subject to a range of assumptions involved in the interpretation of Faraday rotation and synchrotron observations, including the assumption of energy equipartition between cosmic rays and magnetic fields. Beyond semi-quantitative observational and theoretical estimates, the absolute and relative strengths of the two parts of an interstellar magnetic field remain somewhat uncertain. 

The results of the mean-field dynamo theory used in the text are introduced in Appendix~\ref{sec:MFD}; further details and references can be found in \citet{ss21}, among a number of other books and reviews \citep[e.g.,][]{Brandenburg+Subramanian05a}.

\subsubsection{The random (small-scale) field} \label{sec:random}
Random (turbulent) interstellar magnetic fields are produced by the fluctuation (or small-scale) dynamo,  the tangling of the large-scale (or mean) magnetic field by turbulence and compression at random shock fronts.

The fluctuation dynamo amplifies any seed magnetic field on time scales comparable to the turbulent eddy turnover time. Order-of-magnitude estimates suggest that the random field reaches energy equipartition with turbulence after a time of the order of $10\Myr$ \citep[e.g., \citealt{Beck+94}, and sections 6.7 and 13.3 of][]{ss21}. Since this time is short compared to the time scales on which galaxies evolve, we assume that at any given time and position within the galaxy the random field has already saturated, and its rms strength is proportional to $B\eq$, the field strength corresponding to equipartition with the local turbulent kinetic energy density,
\begin{equation}
\label{eq_b}
  b\rms = f_b B\eq\,,
\end{equation}
where
\begin{equation}
\label{eq_Beq}
  B\eq = (4\pi\rho)^{1/2}v\,,
\end{equation}
$\rho$ is the gas density (a function of $r$ and $Z$ as well as of the redshift) and $f_b$ is a constant. While $f_b$ may depend on various parameters of the interstellar turbulence, there is no reliable model at present, so we set it to be constant, chosen based on the existing evidence.

The fluctuation dynamo produces intense magnetic ropes and filaments that do not fill the volume and the rms field strength depends on both the field strength within the filaments and their fractional volume. Most existing simulations of the fluctuation dynamo suggest $f_b\simeq0.5$ due to this mechanism  \citep[\citealt{Brandenburg+Subramanian05a}, see also section~13.3 of][and references therein]{ss21} in incompressible turbulence and even lower in supersonic flows \citep[e.g.,][]{Federrath+14}. However, the simulations are severely limited to modest values of the kinetic and magnetic Reynolds numbers, while $f_b$ can depend on them and is likely to do so. Much stronger, super-equipartition magnetic fields with $f_b>1$ cannot be excluded. For example, this would be the case as long as the fluctuation dynamo produces locally helical magnetic fields, thus diminishing the associated Lorentz force and their back-reaction on the velocity field \citep[see section~4.2 of][for a heuristic model]{BSS93}. The ISM multi-phase structure further complicates the action of the fluctuation dynamo in galaxies \citep{GMLKLS23}.

The tangling of the mean (large-scale) magnetic field by the turbulence (an integral part of the mean-field dynamo action) is thought to produce a volume-filling random magnetic field of a strength comparable to that of the mean-field \citep[section~13.3 of][]{ss21}. Since the mean magnetic field can be stronger than $B\eq$ if the mean-field dynamo is strong, e.g., in the central parts of galaxies where the differential rotation is especially strong \citep[section~13.7 of][and Appendix~\ref{sec:MFD}]{ss21}, this may lead to $f_b>1$.

Compression of magnetic fields at random shock fronts driven by supernovae can produce very strong local random magnetic fields, but their fractional volume is likely to be small; nevertheless, the resulting rms field strength can correspond to $f_b\simeq1$ \citep{ByT87}.

We adopt $f_b=1$, whereas \citetalias{Rodrigues+19} used $0.5$. As the random field is taken to be isotropic, $\mean{b^2_r} = \mean{b^2_\phi} = \mean{b^2_Z} = b\rms^2/3$ for the cylindrical field components.

\subsubsection{The mean (large-scale) field} \label{sec:mean}
\begin{figure*}
\centering
   \includegraphics[width=0.9\textwidth]{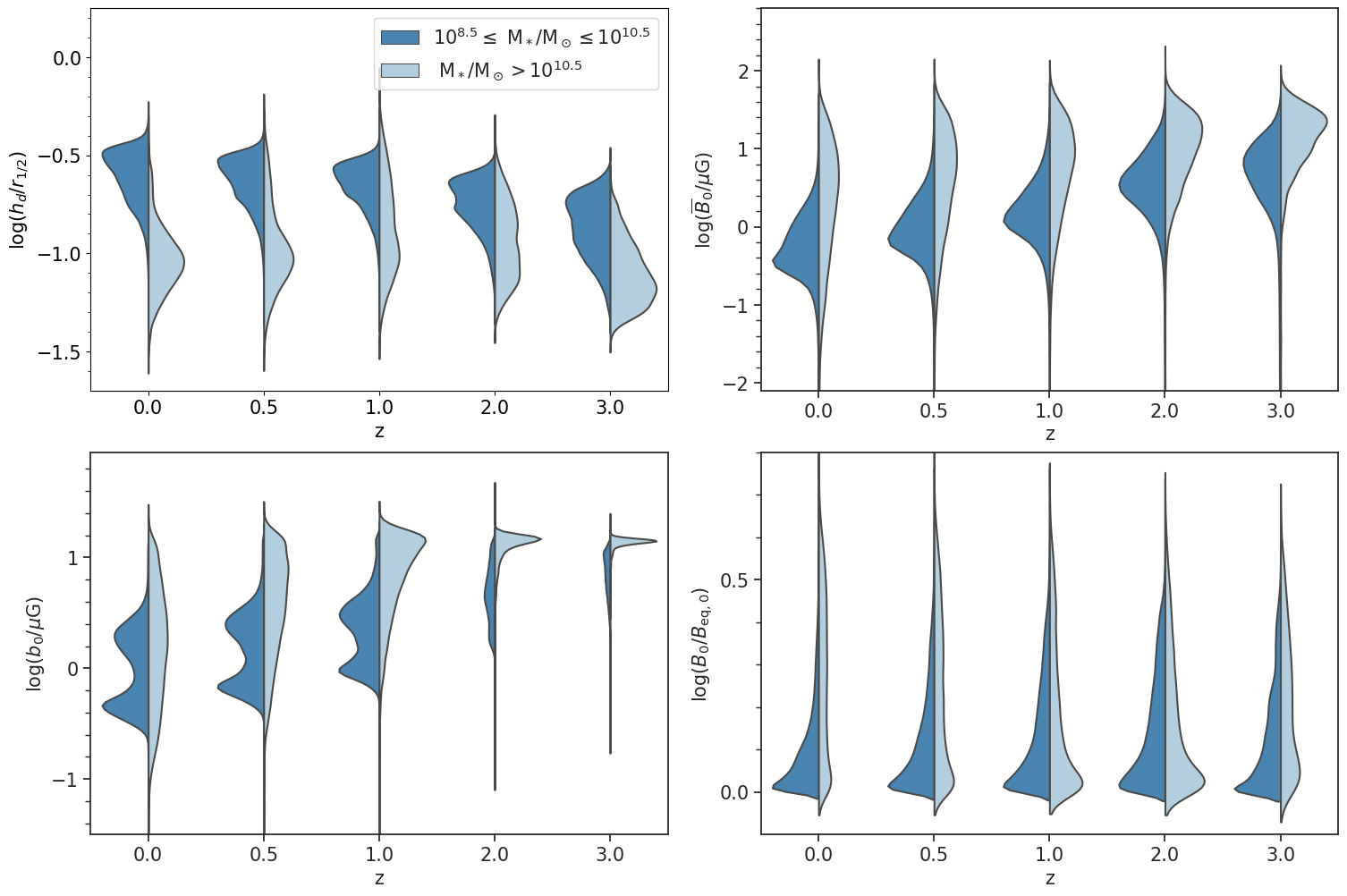}
\caption[]{The probability distributions of the relative gas disc thickness $h\disc(r_{1/2})/r_{1/2}$  (upper left), volume-averaged mean (upper right) and random (lower left) magnetic field strengths, and the ratio of $B_0$ to $B_\text{eq,0}$ (lower right). The magnetic field parameters presented are defined in equations~\eqref{B0} and \eqref{Beq0}. At each redshift, the galaxies are separated into two stellar mass bins defined in the legend.  The area under each curve is proportional to the number of spiral galaxies in the given mass bin at the given redshift.}
\label{fig_violin}
\end{figure*}

The mean-field dynamo, by contrast, can amplify a large-scale magnetic field $\vec{\mB}$ on length scales of the order of a kiloparsec or larger, on time scales that depend on the galactic rotation period, velocity shear due to the differential rotation and the time scale of the turbulent magnetic diffusion across the gas layer. The relevant parameters can be combined into the dynamo number of equation~\eqref{DSA}. The amplification time of the mean field ranges from a fraction of a gigayear in the inner parts of spiral galaxies to a few gigayears in the outer parts. We simulate the evolution of the mean component of the magnetic field in galactic discs by solving numerically the mean induction equation supplemented by a dynamical equation that models the non-linear back-reaction from the Lorentz force on the velocity field that quenches the dynamo to establish a stationary magnetic field. The resulting strength of the large-scale magnetic field is proportional to $B\eq$ and also depends on the dynamo number and other parameters \citep{ss21}. The equations solved and the numerical implementation can be found in \citetalias{Rodrigues+19}, and they rely on the thin-disc approximation in the mean-field dynamo equations applicable for $h/r\lesssim 0.1$ \citep{BSRS87}. The upper left panel of Fig.~\ref{fig_violin} shows that this ratio is less than $0.4$ at the half-mass radius for the vast majority of galaxies, and even smaller for the galaxies in the larger stellar mass range and at a high redshift. To ensure that the solutions of the dynamo equations are computed as fast as required with a large sample of galaxies, we employ well-tested \citep[chapters~11 and 13 of][see also Appendix~\ref{sec:MFD}]{ss21} approximations applicable to a thin disc (\citetalias{Rodrigues+19} and references therein), including the no-$Z$ approximation, which replaces $Z$ (the distance from the mid-plane) derivatives with divisions by the appropriately scaled gas scale height, allowing the equations to depend on only one spatial dimension ($r$). The model parameters are the same as in \citetalias{Rodrigues+19}, with two exceptions. We set $R_\kappa$ \citep{Chamandy+14b} to $1.5$ \citep{Gopalakrishnan+Subramanian23}, whereas \citetalias{Rodrigues+19} used $R_\kappa=1$. The strength of the mean magnetic field is approximately proportional to $R_\kappa$ (see Appendix~\ref{sec:MFD}). The total pressure (the sum of turbulent, magnetic, and cosmic ray pressures) relative to the turbulent pressure is also estimated to be slightly larger than that adopted by \citetalias{Rodrigues+19}.

The initial (seed) magnetic field required to launch the mean-field dynamo is the large-scale component of the random magnetic field produced by the fluctuation dynamo: it does not vanish when the random field is averaged over a finite volume of the galactic disc. This provides an initial kiloparsec-scale random magnetic field of the order of $10^{-9}\G$ in strength, stronger than any plausible primordial magnetic field \citep[\citealt{Ruzmaikin+88,Poezd+93,Bhat+16,Gent+23b}; see chapter~13 of][for a review]{ss21}. We use the same prescriptions as in \citetalias{Rodrigues+19} for initializing the magnetic field at $t=0$ and for preventing the mean field from decaying to very low values when the dynamo is subcritical.

When an evolving galaxy experiences a major merger (where the ratio of the combined stellar and cold gas masses involved exceeds $0.3$), the mean magnetic field is reset to the seed field assuming conservatively that such a merger destroys the large-scale field or distorts it to such an extent that it is so different from an eigenfunction of the dynamo equations that most of it decay rapidly because of the turbulent diffusion and the dynamo starts again with a weak seed magnetic field described above.
\begin{figure*}
\includegraphics[width=0.9\textwidth]{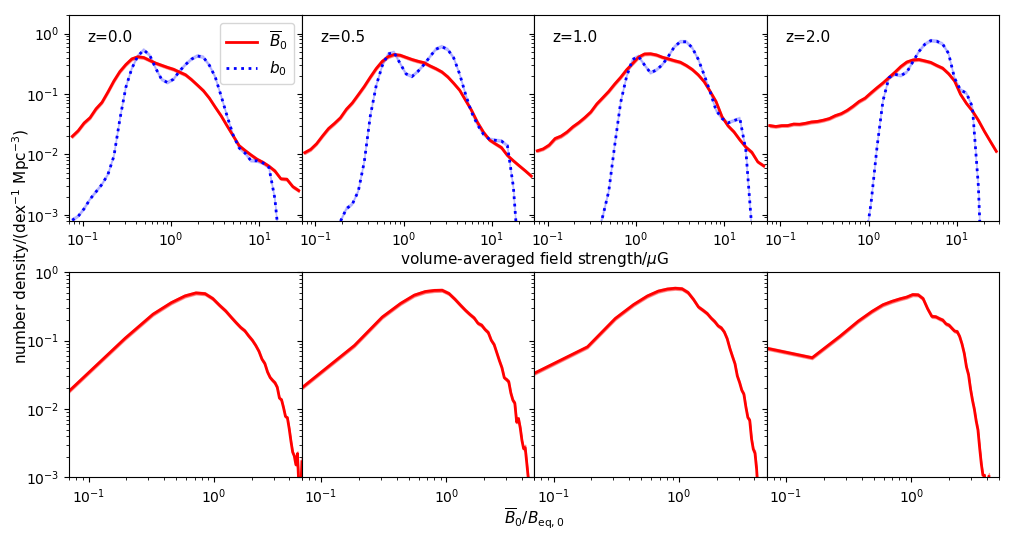}
\caption{The galaxy number density (in comoving coordinates) as a function of various volume-averaged magnetic field properties defined in equations~\eqref{B0} and \eqref{Beq0} for the mean (solid) and random (dotted) fields  (upper panel) and the ratio of $\overline{B}_0$ to $B_\text{eq,0}$ (lower panel) for various redshifts. 
}
\label{num_den_B}
\end{figure*}
\subsubsection{Field distribution across the disc } \label{sss_B_Z_dist}
The procedures described above specify the variation of the random and mean magnetic fields along the galactocentric distance in each galaxy of the sample, but the field distribution across the gas layer (in $Z$) also has to be specified. In our fiducial model, the scale heights, $h_B$, of both the random and the mean magnetic field are assumed to be the same as that of the diffuse gas density $h\disc$. Thus, for example, the distribution in $Z$ of the mean field is adopted as
\begin{equation}
 \vec{\mean{B}}(r,Z) = \vec{\mean{B}}_\text{M}(r) \e^{- |Z|/h_B}\,,
  \label{eq_B_z}
\end{equation}
where $h_B=h\disc$ and $\vec{\mean{B}}_\text{M}(r)$ is the solution of the mean-field dynamo equations produced using  \magnet. 

To provide a simple measure of the magnetic field strength, we use its strength $B_0$ corresponding to the volume-averaged total magnetic energy density, 
\begin{equation}\label{B0}
B_0 = \left[\dfrac{\int_0^{r\disc} B^2 h\disc\, r\, \dd r}{ \int_0^{r\disc} h\disc\, r\,\dd r}\right]^{1/2} ,\;
\end{equation}
for the total magnetic field $\vec{B}$ (where $B\equiv|\vec{B}|$) and similarly $\mean{B}_0$ for its mean part and $b_0$ for the random part. We also introduce a similar quantity for the equipartition magnetic field of equation~\eqref{eq_Beq},
\begin{equation}\label{Beq0}
B_\text{eq,0} = \left[\dfrac{\int_0^{r\disc} B\eq^2 h\disc\, r\, \dd r}{ \int_0^{r\disc} h\disc\, r\,\dd r}\right]^{1/2} .
\end{equation}
\subsection{The synchrotron emission}
\label{secRadioLF}
The mid-radio ($1\text{--}10\GHz$) continuum emission of an SFG consists of the non-thermal (synchrotron) emission due to relativistic (cosmic ray) electrons and the free–free emission from thermal electrons. The radio emission at $1.4\GHz$ is dominated by the synchrotron, with the thermal emission contributing about 10 per cent \citep{Tabatabaei+2017}, and we neglect the thermal contribution to the RLF (see \citealt{Schober+23} for a discussion of the thermal contribution at other frequencies).

We assume that the relativistic electrons have an isotropic energy spectrum
\begin{equation}
N(E)\, \dd E = K_E E^{-s}\, \dd E\,,
\label{eq_CRe}
\end{equation}
with a  factor $K_E$, where $N(E)\,\dd E$ is the number density of electrons with energy between $E$ and $E+\dd E$ and $s$ is the spectral index. The synchrotron emissivity (the energy emitted by the unit volume of the source per unit time and unit frequency interval within the unit solid angle) in a homogeneous magnetic field $\vec{B}$ is then given by \citep[e.g.,][]{GS64,RL79}
\begin{equation}
\label{eq_IB}
\varepsilon= K_E a(s) \frac{e^3}{m\el c^2}
	\left(\frac{3e}{4\pi m\el^3 c^5}\right)^{(s-1)/2}
	B_\perp^{(s+1)/2} 
	\nu^{-(s-1)/2}\,,
\end{equation}
where we use the standard notation for physical constants, $\nu$ is the emission frequency ($=1.4\GHz$ in our case), $B_\perp$ is the total magnetic field strength in the sky plane perpendicular to the line of sight, and 
\begin{equation}
a(s)=\frac{\sqrt3}{4\pi(s+1)}\,\Gamma\left(\frac{3s-1}{12}\right) \Gamma\left(\frac{3s+19}{12}\right),
\end{equation}
where $\Gamma$ is the gamma-function (see chapter~3 of \citealt{ss21} for details). This expression is applicable to a partially ordered magnetic field, $\vec{B}=\vec{\mB}+\vec{b}$ in our case, unless the field is inhomogeneous on very small scales: the synchrotron emission pulse from a relativistic electron is formed over a distance of order $r_B/\gamma\simeq10^9\cm(B/1\mkG)^{-1}$ (where $r_B$ is the electron Larmor radius and $\gamma$ is the particle Lorentz factor), and magnetic fluctuations at larger scales do not affect the applicability of equation~\eqref{eq_IB}. We adopt $s=3$ for the electron spectrum, close to the observed value \citep{BP14,BPA19}. For $s=3$, the statistically averaged value of $\varepsilon$ involves the average of $b_\perp^2$, a quantity predictable theoretically as discussed in Section~\ref{sec:random}.

To derive the synchrotron luminosity, the coefficient $K_E$ in equation~\eqref{eq_CRe} has to be determined. An often-adopted assumption regarding the energy density of cosmic rays (dominated by cosmic ray protons) is their energy equipartition with the magnetic field, 
\begin{equation} \label{eq_crBeq}
  \kappa\cra \int_{E_1}^{E_2} EN(E)\,\dd E=\frac{B^2}{8\pi}\ ,
\end{equation}
where $\kappa\cra$ is the ratio of the energy densities of the relativistic protons and electrons. For $E_2\to\infty$, this yields $K_E = (s-2)B^2 E_1^{s-2}/(8\pi\kappa\cra)$ and, for $s=3$ and $\kappa\cra=100$ \citep[e.g.,][]{BK05}, we have 
\begin{equation}\label{KE}
K_E\simeq  6.3\times10^{-19} \frac{\text{particles}}{\text{cm}^3\,\text{\erg}^{-2}} \left(\frac{B}{1\mkG}\right)^2 \frac{E_1}{1\GeV}\,.
\end{equation}
For $E_1$, we adopt the energy above which the electron spectrum measured in the Solar vicinity has the spectral index $s\approx3$,  
$E_1\approx 8\GeV$ \citep[see fig.~10.10 of][and references therein]{ss21}. Since cosmic rays have a very high diffusivity \citep[e.g.,][]{L94}, we assume that their energy density depends on the spatially averaged magnetic field and use $B^2=\mean{B}^2 +b\rms^2$ in equation~\eqref{KE}.

The assumption of the local energy equipartition between cosmic rays and magnetic fields lacks any general and convincing evidence. Both the observed synchrotron intensity fluctuations in galaxies \citep{SSFBLPT14} and simulations of cosmic ray propagation rather suggest a weak anticorrelation between their energy densities  \citep{TSGSSR22,TSGSS23,TSSS23}. However, there are no readily available, simple alternative approaches to estimate the energy density of cosmic rays in galaxies. For the exploratory study of this paper, we adopt the equipartition assumption but a refinement of our model would include an estimate based on the SFR and a cosmic ray propagation model.

Using the large-scale and small-scale magnetic fields computed for each galaxy, we derive $B_\perp$ and obtain the synchrotron luminosity assuming that the galaxies in the sample have random orientations with respect to the line of sight, (we use the uniform distribution of $\cos i$, where $i$ is the inclination angle of a galaxy to the line of sight), 
\begin{align}
L&=\int \dd \Omega \int_{V} \varepsilon(\vec{r})\,\dd ^3\vec{r} \nonumber\\
&=8\pi \int_0^{r\disc} r\,\dd r \int_0^{2\pi} \dd \phi \int_{0}^{h\disc} \dd Z \, \varepsilon(r,\phi, Z)\,,
  \label{eq_syn_Lum}
\end{align}
where $\Omega$ is the solid angle. The method for computing the integral in equation~\eqref{eq_syn_Lum} is described in Appendix~\ref{appen_syn_lum_1D}.
\section{Magnetic fields in evolving galaxies}\label{MFiEG}
The top panel of Fig.~\ref{num_den_B} shows the number density of galaxies in the fiducial model as a function of the volume-averaged field strengths of the mean and random magnetic fields,  $\mB_0$ and $b_0$ defined in equation~\eqref{B0} in each galaxy at various redshifts. 
At $z=0$, galaxies with $1\mkG \lesssim b_0 \lesssim 5\mkG$ are more common than galaxies with $\overline{B}_0$ in the same range, while galaxies with $\overline{B}_0<0.4\mkG$ or $\overline{B}_0>10\mkG$ are far more common than those with $b_0$ in that range. With redshift, these limits shift to higher values of magnetic field strength. At lower redshifts, the number density of galaxies as a function of $b_0$ has two distinct maxima. The distribution of $\overline{B}_0$ is wider than that of $b_0$ because $b\rms$ is directly related to $B_\mathrm{eq}$ through equation~\eqref{eq_Beq}, whereas $\overline{B}$ depends on other galactic parameters as well. The lower panel of Fig.~\ref{num_den_B} shows the number density of galaxies as a function of $\mB_0/B_\mathrm{eq,0}$ and since $b\rms=B_\mathrm{eq}$ for $f_b=1$ (Section~\ref{sec:random}) and hence $b_0=B_\text{eq,0}$, it is clear that most galaxies in our model have a stronger random magnetic field compared to the mean field. As a larger volume-averaged field strength typically results in a higher synchrotron luminosity, magnetic fields in radio-bright galaxies are predominantly random, especially at high redshift (see also Section~\ref{sec_radioLF}). 

Figure~\ref{fig_violin} shows the evolution of the probability density of the mean and random magnetic field components in the redshift range $0<z<3$ with galaxies grouped into two stellar mass bins,  $8.5 \leq \log(M_\ast/\!\msun) < 10.5$ and $\log(M_\ast/\!\msun) \ge 10.5$. The probability density is normalized such that the total probability over all redshifts is unity in each mass bin, separately. This allows us to visualize the relative number of galaxies with a given magnetic field both at each redshift and between different redshifts. The highest mass bin contains about $3\%$ of the total number of galaxies in our sample, and the number of galaxies in a given mass bin changes with $z$ owing to star formation, mergers,  etc. The probability distribution of $\overline{B}_0$ has a single peak, with most galaxies having $\overline{B}_0>0.1\mkG$ at all redshifts. This is different from the results of \citetalias[][their fig.~6]{Rodrigues+19}, where a significant fraction of galaxies have very weak magnetic fields of the order of $10^{-4}\mkG$ (the strength of the seed field) because the mean-field dynamo is subcritical in some galaxies, particularly at higher redshifts. The mean-field dynamo is significantly stronger in our fiducial model because of the higher turbulent speed $v$. Stronger turbulent flows promote the mean-field amplification despite the increased turbulent diffusion because the dynamo number in the kinematic regime scales as $h\disc^2/v^2$ (see Appendix~\ref{sec:MFD}) and $h_\mathrm{d}$  increases with $v$ faster than linearly. Indeed, under hydrostatic equilibrium, the disc scale height is proportional to the gas pressure, which includes the turbulent contribution $\rho v^2$, so $h\disc$ increases with $v$ approximately as $v^2$, and therefore, the dynamo number increases with $v$.
Our sample includes massive galaxies with a volume-averaged mean field strength as high as $100\mkG$, even at high redshifts. Since $\overline{B}_0$ is a volume-averaged quantity, the local mean magnetic field strength can be significantly higher. This is consistent with the observational results of \citet{Geach+2023} where a density-weighted ordered field as strong as $500\mkG$ is estimated for the dense, dust-rich regions of a galaxy at $z=2.6$. 

\begin{figure*}
\includegraphics[width=0.8\textwidth]{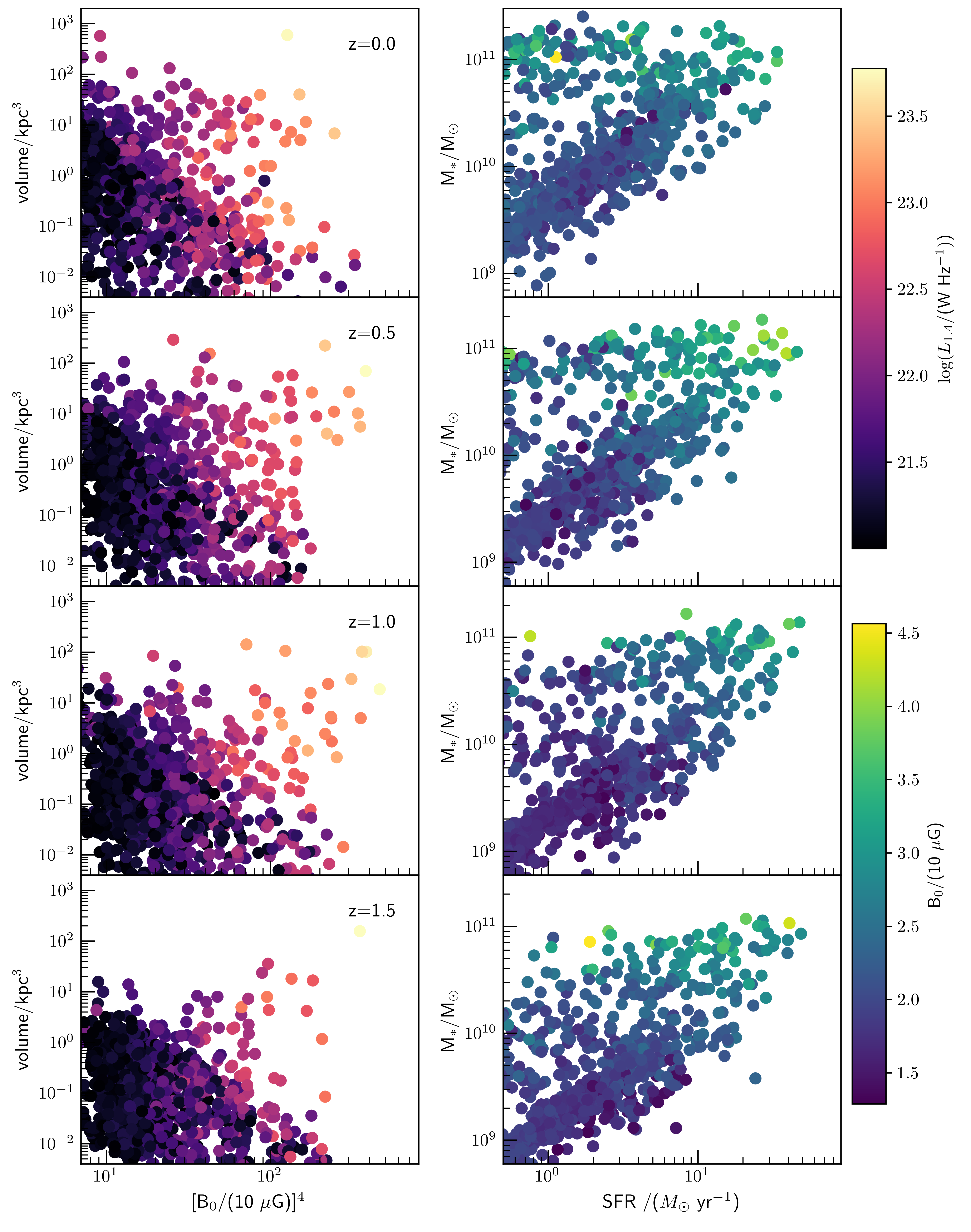}
\caption{Left: the scatter plots of the colour-coded radio luminosity $\log(L_{1.4})$ of $10^3$ randomly selected galaxies with $L_{1.4} \geq 10^{21} \W\Hz^{-1}$ at $z=0.0, 0.5, 1.0$ and $1.5$ versus the fourth power of the volume-averaged magnetic field strength and the volume of the emission region (disc volume). Right: the colour-coded volume-averaged magnetic field strength of galaxies shown in the left-hand panels versus their star formation rate and stellar mass. }
\label{L1.4_z_evol}
\end{figure*}

Figures~\ref{num_den_B} and \ref{fig_violin} also show that the mean magnetic field across a sample of galaxies decreases with time because of the depletion of the interstellar gas (although individual galaxies may have monotonically increasing field strength depending on their formation history), which is consistent with the results of \citetalias{Rodrigues+19} and \citet{Rodrigues+15}. We also find, as \citetalias{Rodrigues+19} do, that the mean field strength shows a stronger scatter for galaxies with higher stellar masses. However, a new feature of the present model is that galaxies with larger stellar masses have stronger mean magnetic fields, on average, whereas in \citetalias{Rodrigues+19} there was less of a distinction. Galaxies with higher stellar mass generally have higher SFR, as seen in the right-hand panels of Fig.~\ref{L1.4_z_evol}, where we show the scatter plot of $M_\ast$ versus SFR, with data points coloured according to the volume-averaged total magnetic field strength. The correlation between the SFR and $M_\ast$ (the so-called galaxy main sequence) is consistent with several observational studies and galaxy formation models 
\citep{Dave2008, Davies+2017,  Bauer+2011, Whitaker+2014, MartinS+2015, Katsianis+2016, Cowley:2017}. In our fiducial model, $v$ increases with SFR. The strength of the saturated mean field increases with $v$ because of the increased dynamo number and the enhancement of $B\eq$ of equation~\eqref{eq_Beq}. This explains the increase of $\mean{B}_0$ with the stellar mass seen in Fig.~\ref{fig_violin}. The increase of the total magnetic field strength $B_0$ with the stellar mass, evident in Fig.~\ref{L1.4_z_evol}, is explained by the increase  of both $\mean{B}_0$ and $b_0$ ($\equiv B_\mathrm{eq,0}$) with $M_\ast$. 

Figure~\ref{fig_violin} also shows the probability density of the volume-averaged random magnetic field strength $b_0$. Similarly to $\overline{B}_0$, galaxies of a higher stellar mass have larger $b_0$. This can be explained by the higher SFR of such galaxies, hence higher $v$ according to equation~\eqref{eq_v0} and larger $B\eq$ according to equation~\eqref{eq_Beq} (we have confirmed that the correlation of $\rho$ with $M_*$ is very weak). The distribution of $b_0$ is bimodal for low-mass galaxies, but not for galaxies of a high mass, for reasons not yet quite clear to us. 

The lower panel of Fig.~\ref{fig_violin}  shows the distribution of the ratio of the characteristic field strengths $B_0$ and $B_\text{eq,0}$ introduced in equations~\eqref{B0} and \eqref{Beq0}. The ratio $B_0/B_\text{eq,0}$ tells us how the volume-averaged magnetic energy density of a galaxy $B_0^2/8\pi$ compares to its averaged turbulent energy density $B_\text{eq,0}^2/8\pi$ and how close are the dynamos  to the saturated state. The random magnetic field saturates at $B\eq$ in our model while the mean field can be amplified to significantly larger strengths.

The ratio $B_0/B_\text{eq,0}$ is sensitive to the distribution of the magnetic field within the disc. The energy density of the turbulence, given by $B\eq^2/8\pi$ from equation~\eqref{eq_Beq},  decreases with the galactocentric radius $r$ because of the reduction in the gas density $\rho$ alone as we assume that the turbulent speed $v$ is independent of $r$. The strength of the mean magnetic field $\mean{B}$, while generally decreasing with $r$, can have a different radial distribution as it depends, in particular, on the angular velocity and shear rate of the galactic rotation. As an illustration, consider the radial dependencies $\mean{B}= \tilde{B} \exp(-r/r_B)$, $B\eq=\tilde{B}\eq\exp(-r/r\eq)$ and $h\disc=\tilde{h}\disc \e^{r/r_h}$ with certain length scales $r_B$, $r\eq$ and $r_h$ for $0\leq r <\infty$ (and the variables with the tilde denoting the central values, those at $r=0$), so that 
\begin{equation}\label{B0Beq0}
\frac{\mean{B}_0}{B_\text{eq,0}}=\frac{\tilde{B}}{\tilde{B}\eq}\, \left|\frac{2r_h/r\eq-1}{2r_h/r_B-1}\right|\,.
\end{equation}
At all redshifts, the probability distributions of  $\mean{B}_0/B_\text{eq,0}$ have extended tails, especially in the case of massive galaxies, where this ratio exceeds unity (up to three) suggesting that most galaxies have $\tilde{B}/\tilde{B}\eq\gtrsim 1$ and/or $r_B\gtrsim r\eq$. The mean magnetic field strength likely exceeds $B\eq$ in the central parts of galaxies where the dynamo is the strongest. A very slow decrease in the strength of the mean magnetic field with $r$, $r_B\gg r\eq$, has been observed in nearby galaxies \citep[section~4.2 of][and references therein]{Beck15}. A large radial extent of magnetised region in spiral galaxies is explained by the slow decrease of the dynamo number with $r$ because of the disc flaring which compensates for the relatively slow decrease of the rotation rate in a flat rotation curve, as can be seen from equation~\eqref{DOH}. The radial extent of the magnetised region is controlled by the decrease of the gas density and $B\eq$ with $r$.

\section{The radio luminosity function}
\label{sec_radioLF}
\begin{figure*}
\includegraphics[width=0.8\textwidth]{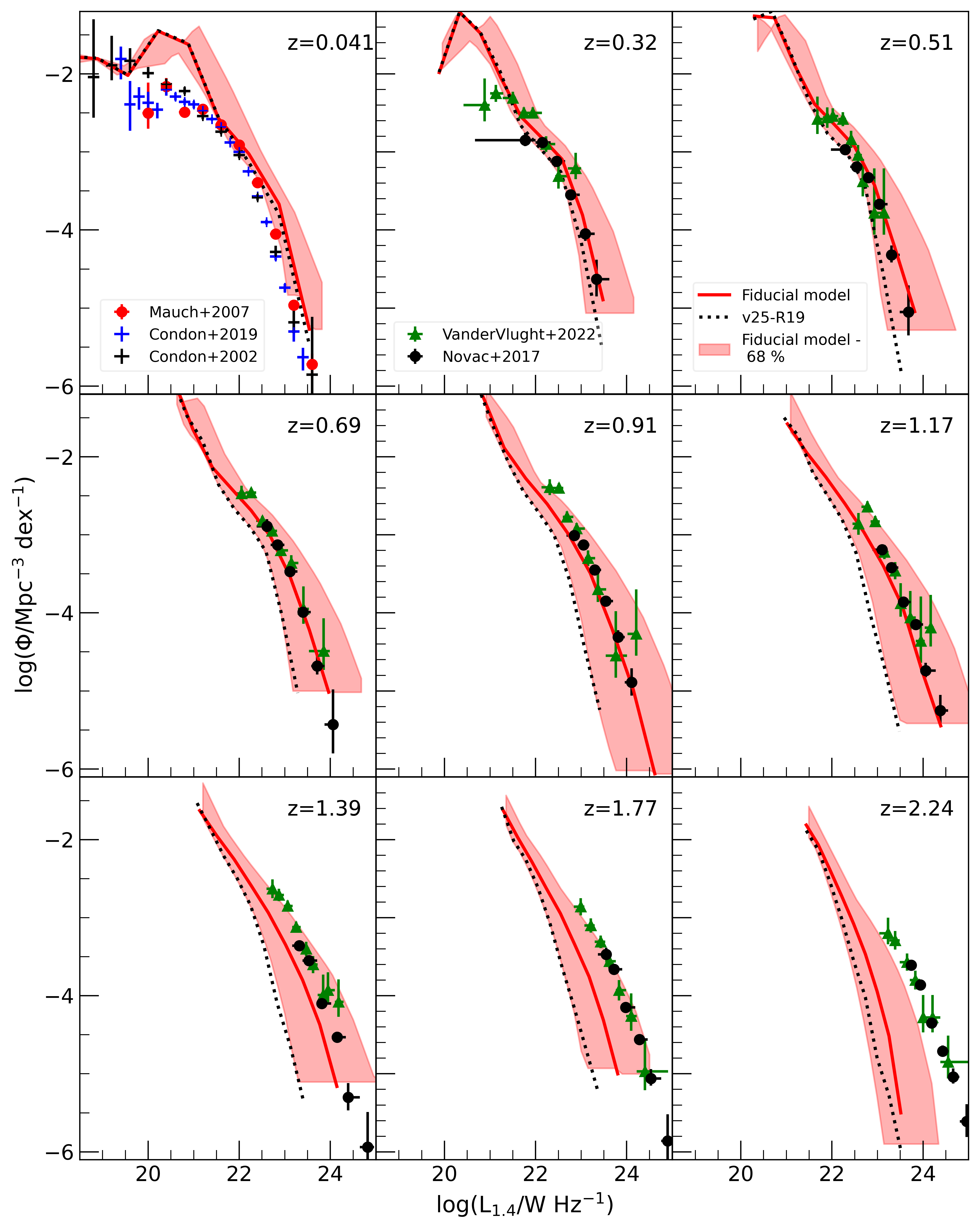}
\caption{The 1.4-GHz radio luminosity function of star-forming galaxies at various redshifts $z$ specified within each frame. The median outcome of the fiducial model is shown as a red solid line with the shaded region representing the 16--84 percentile range of the turbulent speed shown in Fig.~\ref{fig_turbv0}. The $v$25-R19 model (with the constant turbulent speed $v=25\kms$) is represented with the dotted black line. The observational data are  taken from \citet{condon+2002} (black crosses), \citet{condon+2019} (blue crosses), \citet{Mauch_Sadler_2007} (red circles), \citet{Novak_2017} (black circles) and \citet{vanderVlugt_2022} (green triangles). }
\label{RLFfid}
\end{figure*}
Figure~\ref{RLFfid} shows the variation with the redshift of the RLF $\Phi(L_{1.4})$ of SFGs at the rest-frame frequency $1.4\GHz$, where $\Phi(L_{1.4})$ is the number of galaxies per unit comoving volume per decade in luminosity, at luminosity $L_{1.4}$. For comparison, we note that the Milky Way luminosity at $1.4\GHz$ is $L_0=3\times 10^{21}\W\Hz^{-1}$ \citep{B84}. The predictions are computed at the redshifts (shown on each panel) closest to the median of the redshift bins used for RLF measurements by \cite{condon+2002} and \cite{vanderVlugt_2022} where simulation snapshots are available. We show the RLFs obtained from two models described in Section~\ref{ism_turbulence}: the fiducial model (where the turbulent speed $v$ depends on the SFR) and the model $v$25-R19 where $v=25\kms$ in all galaxies. The observed luminosity functions are from \citet{condon+2002, condon+2019} and \citet{Mauch_Sadler_2007} for the local Universe and \citet{Novak_2017} and \citet{vanderVlugt_2022} for the higher redshifts. 

The agreement of the fiducial model with the data is quite satisfactory, if not perfect, over a wide redshift range. In particular, for $z \lesssim 1.2$, the high-luminosity end of the RLF is reproduced remarkably well. Although the median luminosity is under-predicted at higher redshifts for all values of $L_{1.4}$,  the model and observations still marginally agree for $z \lesssim 2$, including the changes in the slope around $L_{1.4}= 10^{23} \W\Hz^{-1}$ for $z \lesssim 0.5$. Moreover, when extrapolated to higher luminosities, the median outcome of the fiducial model is in agreement with the data point at the highest luminosity at small $z$ (our galaxy sample is not large enough to probe such bright, rare galaxies at certain redshifts, including $z \approx 0$).  However, the number density of galaxies in the luminosity range $20.5 \leq \log (L_{1.4}/\W\Hz^{-1})\leq 21.5$ in the local Universe is significantly higher than what is observed. This local maximum at lower luminosities is sensitive to the poorly constrained ratio of the mean and random magnetic fields (Section~\ref{sec_bvsB}) and might be adjusted by the fine-tuning of our model, which we avoid. Also, for  $z > 2$, the model number densities are systematically and significantly smaller than those observed, and the discrepancy between the predictions and the data increases with redshift. 
\begin{figure*}
\includegraphics[width=0.8\textwidth]{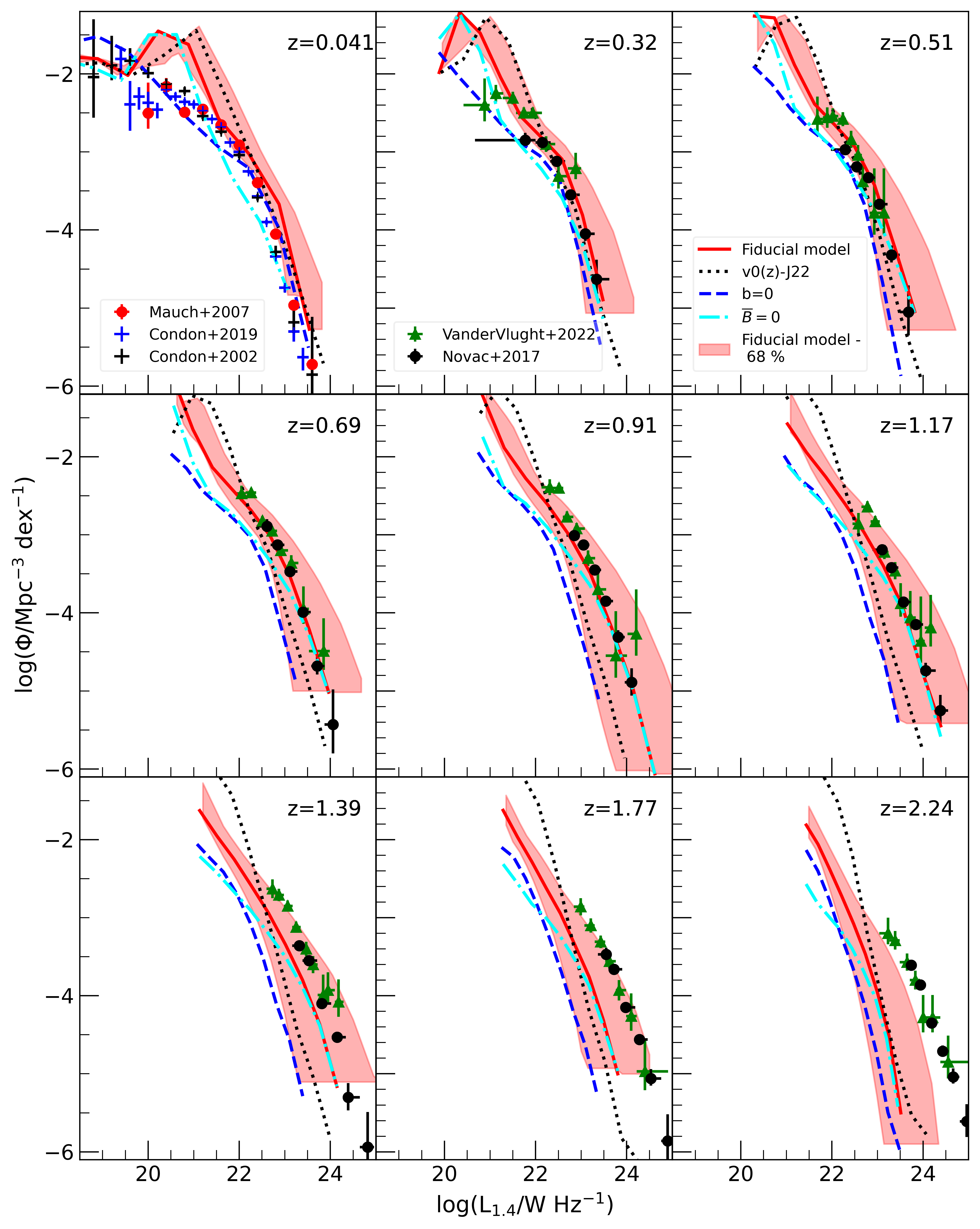}
\caption{As Fig.~\ref{RLFfid}, but the fiducial model compared with the $v(z)$-J23 model (black dotted) where the turbulent speed is an explicit function of the redshift. The dashed blue and dash-dotted cyan lines show the contributions to the synchrotron luminosity due solely to the mean and random magnetic fields, respectively.} 
\label{RLFother}
\end{figure*}
\subsection{The role of the turbulent speed}
\label{TRofTS}
The turbulent speed $v$ is among the model parameters that affect directly the RLF because the strength of the magnetic fields on both large and small scales increases in proportion to the turbulent kinetic energy via their dependence on $B\eq$ through equation~\eqref{eq_b}, and the large-scale field strength has additional dependence on $v$ via the $\alpha$ effect and turbulent diffusion. The models $v$25-R19 and $v(z)$-J23 are designed to explore the role of this parameter.

Apart from the fiducial model results, Fig.~\ref{RLFfid} also shows the prediction of the $v$25-R19 model  (dotted black) which has the constant turbulent speed $v=25\kms$ in all galaxies, in contrast to the fiducial model where  $v$ is larger in galaxies with $\SFR\ge3\,\msun\yr^{-1}$ (Section~\ref{ism_turbulence}). The two models differ little in the predicted RLFs at low luminosities, especially at low redshifts where  $\SFR<3\msun\yr^{-1}$  and hence $v=25\kms$ in most galaxies. However, the increase in the SFR up to $z\simeq2$ makes the $v$25-R19 model predictions differ from those of the fiducial model, and they become less and less adequate with increasing $z$, severely under-predicting the number density of galaxies at the high luminosity end. Even if the overall luminosity were increased by a factor by adjusting model parameters (as discussed in Section~\ref{Disc}), the shape of the $v$25-R19 RLF does not match the data as well as the fiducial model. 

Figure~\ref{RLFother} compares the fiducial model with the model $v(z)$-J23 (dotted black), where $v$ is an explicit function of the redshift taken from the bottom panel of Fig.~\ref{fig_turbv0}. Neither the overall shape nor the magnitude of the resulting RLFs agree with the observations: this model is inferior to the fiducial one. 

In fact, neither the fiducial nor the $v(z)$-J23 model is fully justified because model~$v(z)$-J23 does not capture the SFR-dependence of turbulence at constant $z$, while the fiducial model does not allow for such effects as the gravitational instability \citep{Krumholz_18} and gas accretion and mergers \citep{Ginzburg_Dekel+22} which are likely to result in an additional dependence of the turbulent intensity on redshift independently of the SFR. A more realistic model for turbulence will be among the refinements of our model to follow. 

\subsection{Relative importance of the random and mean magnetic fields}
\label{sec_bvsB}
Under the assumption of energy equipartition between cosmic rays and magnetic field adopted here, the contributions of the mean and 
random magnetic fields to the RLF are not additive because the radio luminosity includes the integrals of $(\vec{\mB}_\perp\cdot\vec{b}_\perp)$, $\mB^2 b\rms^2$ and $\mB_\perp^2 b\rms^2$. Therefore, to assess the significance of the mean and random magnetic fields, we consider results obtained when one of them is ignored. In Fig.~\ref{RLFother}, we show the predictions with only the random field (dash-dotted cyan line) or the mean field (dashed blue line) retained. Both random and mean fields are clearly important. However, the RLF that relies solely on the mean magnetic field decreases with the luminosity too rapidly and by $z\approx1$ the contribution from the random field dominates that of the mean-field at least at those high luminosities where observational data are available. On the other hand, at $z\approx 0$, we obtain a significantly better fit to the data by neglecting the random field and including only the mean field, particularly at low luminosities where the fiducial model over-predicts the RLF. 

We stress that the relation between the contributions of the random and mean magnetic field components to the RLF depends on poorly constrained parameters like $f_B$ for the random field and the magnitude of the mean helicity of interstellar turbulence for the mean field. Since the radio luminosity scales as the fourth power of the magnetic field (Section~\ref{sec_radioLF}), the relative influence of the random and mean magnetic fields is rather sensitive to such parameters. Clearly, the same magnitude of the luminosity function could be obtained by changing the parameter values so as to boost one field component and reduce the other. Therefore, the relative importance of mean and random field is not well-constrained by the model and for realistic parameter values, neither can be neglected. 

The current interpretations of the observations of the synchrotron emission in nearby galaxies suggest that the energy density of the random magnetic field exceeds that of the large-scale field by a factor of three or even larger \citep{Beck+19}. However, the interpretations rely on the assumption of the local energy equipartition between cosmic rays and magnetic fields, and so are model-dependent. 

To conclude, there are several plausible reasons for the apparent complexity in the relative roles of the mean and random magnetic fields at different redshifts. It is likely that a better model for cosmic rays is necessary (e.g. because their number density might be more sensitive to the mean magnetic field than to its random part). Another reason could be our limited understanding of the interaction of the fluctuation and mean-field dynamos \citep[chapter 8 of][]{ss21}. Both types of dynamo action are sensitive to the multi-phase ISM structure but in a poorly understood manner while it is clear that the ISM structure depends on both the SFR and redshift. Remarkably, the RLF can shed light on rather subtle aspects of both the ISM structure and galactic dynamos. The relative importance of mean and random fields thus deserves careful analysis beyond the scope of this paper.
\subsection{The redshift evolution of the RLF}
\label{sec_LF_z_evol}
The redshift evolution of the observed RLF  for bright star-forming galaxies ($L_{1.4}\gtrsim10^{21}\W\Hz^{-1}$) is discussed by \citet{Novak_2017} and \citet{vanderVlugt_2022,vanderVlugt_2023}; their observational data are shown in Figs~\ref{RLFfid} and \ref{RLFother} together with our results. Both the fiducial model (where the turbulent speed increases with the SFR) and the observations agree that the range of the radio luminosities extends to larger $L_{1.4}$ as $z$ increases. The galaxy luminosity functions observed in far infrared \citep{Gruppioni+2013, koprowski+2017, Lim+2020} and K-band \citep{cirasuolo_2010_Kband, mortlock_2017_Kband} evolve similarly with the redshift. However, the RLF shows a slightly stronger evolution compared to that in the infrared as the infrared-to-radio luminosity ratio is observed to decrease mildly with increasing redshift \citep{delhaize+2017_IR_radio,riveraC+2017_IR_radio, magnelli+2015_IR_radio, ivison+2010_IR_radio}. 

Both the size of the emission region and the strength of the magnetic field contribute to the increase of the synchrotron luminosity with $z$. This is illustrated in the left-hand panels of Fig.~\ref{L1.4_z_evol}, which presents the colour-gradient scatter plot of the radio luminosity of a thousand randomly selected galaxies with $L_{1.4} \geq 10^{21} \,$WHz$^{-1}$ from our sample at different redshifts as a function of their disc volume and $B^4_0$ (as $L_{1.4} \propto B^4$ under the assumption of energy equipartition between cosmic rays and magnetic fields). It is clear that the brightest galaxies typically have \textit{both} large emitting volume and volume-averaged magnetic field strength. For example, the volume-averaged magnetic field strength of the brightest galaxies at $z=1.5$ is about a factor of two larger than at $z=0$ 

Certain factors that contribute to stronger magnetic fields at high redshifts are clarified in the right-hand panels of Fig.~\ref{L1.4_z_evol}. Galaxies with larger $M_\ast$ tend to contain stronger magnetic fields, for reasons explained in Section~\ref{MFiEG}. The SFR density has a maximum at $z \simeq  1.0$ (Fig.~\ref{figSFRoffset}), indicating high SFR in galaxies at high redshifts. This results in a higher turbulent speed in galaxies via equation~\ref{eq_v0} leading to a systematic increase of the magnetic field strength with $z$ at $0\lesssim z\lesssim1.5$. 

Thus, our fiducial model finds a shift to higher luminosities with increasing $z$, in agreement with the observations up to  $z\approx1.8$. However, as mentioned at the beginning of Section~\ref{sec_radioLF}, the $z$-dependent shift in the model RLF is weaker than that inferred from observations. The median predictions of the model seem to underestimate the observed RLF at $z\gtrsim1$, and the difference between them increases with increasing $z$, and at $z\gtrsim2$ they clearly disagree for reasons discussed in the next section.
\section{Discussion}
\label{Disc}
 Our goal in this paper is, firstly, to explore the applicability of the theory and models of galactic magnetic fields to young and evolving galaxies and, secondly, to identify the most important galactic properties and parameters that can be deduced and understood with the  help of their radio luminosity function.  The synthetic RLFs of star-forming galaxies obtained from their modelled synchrotron emission at $1.4\GHz$ are in broad agreement with the available observations, reproducing the number density of galaxies in the luminosity range $10^{19}\lesssim L_{1.4}\lesssim10^{25}\W\Hz^{-1}$ \citep[$3\times10^{-3}$ to $3\times10^3$ in terms of the Milky Way luminosity at $1.4\GHz$ --][]{B84} in the redshift range  $0\lesssim z\lesssim2$, within the uncertainties in the turbulent speed discussed in Section ~\ref{ism_turbulence}. Moreover, the form (e.g., changes in the slope) of the RLF agrees well with what is observed. We do not consider $z \gtrsim 2$, because the synchrotron luminosity predicted by the model becomes significantly lower than what is observed, continuing the trend with $z$ already 
 visible in Fig.~\ref{RLFfid}. 

We have made no persistent attempt to achieve any better agreement with the observational results even though this would be possible given that many properties of the ISM in evolving galaxies leave much freedom for adjustment. We avoid making heuristic assumptions that do not have explicit observational or theoretical justification.

We have identified the turbulent speed $v$ as one of the predominant factors affecting the RLF. Our results firmly indicate that $v$ must be an increasing function of the star formation rate for the SFR exceeding a certain threshold (here adopted as $\SFR_0=3\msun\yr^{-1}$), and the model with $v=\const$ ($v$25-R19) cannot explain the increases in the galactic luminosity with $z$ even though its results are close to those with the SFR-dependent turbulent speed (the fiducial model) at $z\approx0$ (Fig.~\ref{RLFfid}).

We also find that the shape and redshift-dependence of the RLF cannot be reproduced by making $v$ depend explicitly on $z$ instead of on the SFR. This is elucidated by the model $v(z)$-J23, which predicts a decrease of the RLF with luminosity that is too steep, failing badly to reproduce observations.

We appreciate, however, that the variation of the turbulent speed with the SFR adopted in the fiducial model may just serve as a proxy for other effects discussed below.
\subsection{Interstellar magnetic fields}
\label{IMaF}
The energy density of the random magnetic field  is proportional to the energy density of the turbulence with the proportionality factor $f_b^2$ of equation~\eqref{eq_b} only known to be of order unity. Adopting $f_b=1$, we find a reasonable agreement between the predicted RLF and observations at the redshifts $z\lesssim1.2$ (see Fig.~\ref{RLFfid}). Our knowledge of the strength and structure of turbulent magnetic fields largely relies on numerical simulations severely limited to modest Reynolds numbers. For example, adopting $f_b=1.5$ improves significantly the agreement at higher redshifts. Values of $f_b$ exceeding unity can be due to a stronger magnetic helicity since partially helical fields are close to the force-free state, so their back-reaction on the flow is reduced, which allows them to have a higher energy density. Thus, our model might fit the observations much better, had we assumed that $f_b$ increases slightly with $z$. Moreover, $f_b$ can well depend on various galactic parameters and be sensitive to the multi-phase ISM structure rather than be a constant or a simple function of $z$. Our results highlight the need to understand better how random magnetic fields are maintained in galaxies. The theory of the mean magnetic fields is developed better although many aspects of the action of the mean-field dynamo in the multi-phase, evolving ISM remain speculative. Since the energy density of both the turbulent and large-scale magnetic fields depends on the turbulent energy density, the gas density and turbulent speed are identified as major factors affecting the RLF. However, the mean magnetic field has a more complicated dependence on the galactic parameters than its random part, including the differential rotation and the gaseous disc thickness in addition to the turbulent energy density. The energy density of the mean magnetic field (unlike the random field in our model) can exceed the turbulent energy density. As a result, the total magnetic energy density in a significant fraction of galaxies (especially massive ones) exceeds the turbulent energy density \citep[see also fig.~6 of][]{Rodrigues+19}. The RLF in polarised emission would help to isolate the contribution of the mean magnetic field to the synchrotron luminosity and thus assess rather subtle properties of evolving galaxies. Our predictions for the polarised RLF will be published elsewhere.
\subsection{Galactic structure}
\label{GS}
The predictions of our fiducial model, which assumes that the structure of a galaxy is broadly similar to a spiral galaxy in the local Universe, are reasonable until $z\approx2$, the redshift at which galaxies appear to develop persistent thin discs dominated by rotation similar to those at $z\approx0$ \citep[][and references therein]{Ginzburg_Dekel+22, Jimenez_Lagos+23}. Our models implicitly assume that the disc is formed instantaneously whereas the history of the disc formation can affect the outcome of the evolution in the strongly nonlinear galactic systems. Certain effects of enhanced star formation on the galactic structure and environment, such as outflows (fountains and winds) and the emergence of radio haloes, are not captured by our model which only includes magnetic fields in galactic discs. Such effects can affect high-redshift galaxies more profoundly and may help to explain why our model does worse at matching observations as $z$ increases. For example, galaxies with a high SFR may have radio haloes with vertical extent 5--$10\kpc$ and magnetic field strength comparable to that in the disc. Thus, future improvements to the magnetic field model should include the radio halo and its contribution to the synchrotron luminosity.
\subsection{Cosmic rays}\label{CR}
To obtain the number density of cosmic ray electrons, we have adopted a widely used (albeit poorly justified) assumption of energy equipartition between cosmic rays and magnetic fields. However, the energy density of cosmic rays (including the number density of relativistic electrons) is likely to be an increasing function of the supernova rate directly related to the SFR. We shall explore elsewhere an alternative model where the abundance of cosmic rays is controlled by their sources and propagation in the evolving magnetic field.

The predicted RLF is also sensitive to the slope $s$ and amplitude K$_E$ of the energy spectrum of cosmic ray electrons given by equation~\eqref{eq_CRe}. Under the assumption of energy equipartition between cosmic rays and magnetic field, changes to the value of $s$ also affect $K_E$ (in proportion to $s-2$) and hence the predicted luminosity of all galaxies, shifting the RLF along the horizontal ($\log L$) axis. The shape of the RLF also can be mildly modified as the synchrotron emissivity is proportional to $B_\perp^{(s+1)/2}$ (see equation~\ref{eq_IB}). Since the emissivity changes as $\nu^{-(s-1)/2}$, multi-wavelength radio observations can constrain $s$ and resolve the degeneracy between K$_E$ and $s$.

The estimate of the synchrotron emissivity of equation~\eqref{eq_IB} needs to be refined at high redshifts because the energy losses of relativistic electrons to the inverse Compton scattering off the CMB photons increase as $(1+z)^4$. 
The ratio of the rates at which a relativistic electron loses energy to the inverse Compton scattering and synchrotron is given by $8\pi\, w_\text{CMB}/B^2$, where $w_\text{CMB}$ is the CMB energy density \citep[section~3.1.4 of][]{ss21}. For electrons with the Lorentz factor $\gamma$, the frequency of the CMB photons is boosted by the factor $\gamma^2$, so $1\GeV$ electrons emit in the X-ray range. \citet{ScBe13} suggest that the inverse Compton losses dominate over the synchrotron emission at $z\gtrsim 2$ if the typical strength of the galactic magnetic fields $B$ does not increase with $z$ but at a larger $z$ if $B$ increases with the redshift. The effect of the inverse Compton losses on our results at $z\leq2$ is likely to be only modest, especially for the radio-bright galaxies.
\subsection{Interstellar turbulence and galactic outflows}\label{ITGO}
It appears that the mechanisms of magnetic field generation, and the ISM structure on which they rely, experience a significant change at $z\simeq2$. This is suggested, in particular, by the increase in the gas velocity dispersion with the SFR (Fig.~\ref{fig_turbv0}) above the sound speeds in the diffuse ISM phases, $c\sound\simeq10\kms$ in the warm gas and $c\sound\simeq130\kms$ in the hot phase. Meanwhile, simulations of the supernova-driven multi-phase ISM show that the fractional volume of the hot gas at the galactic mid-plane increases only slightly (as $\SFR^{0.36}$ assuming that the supernova rate is proportional to the SFR) from about 0.2 to 0.28 when the supernova rate increases by a factor of four in comparison with the Milky Way rate and does not exceed 0.5 when the supernova rate is 16 times the local rate, while the abundance of the cooler phases reduces as the supernova rate increases \citep[][and references therein]{dAB04,BdA21}. The main consequence of the increase in the SFR is a more vigorous outflow of the hot gas in the form of the galactic fountain or wind. Although the off-planar gas is hot and highly ionised, significant amounts of neutral hydrogen are present, entrained from the disc, or cooled down \textit{in situ}). Its \ion{H}{i} velocity dispersion is about $60\kms$ in the Milky Way \citep[table~1 of][]{K03} while the velocity dispersion of the hot ionised gas is $75\kms$ \citep[section~8.3 of][]{SSL97}. A review of relevant observations and further references can be found in section~10.2.2 of \citet{ss21}. Meanwhile, the turbulent velocity in the disc hardly changes as the supernova rate varies by a factor 512\luke{,} remaining close to the sound speed in the warm gas while the \ion{H}{i} line width is larger than the (one-dimensional) turbulent velocity dispersion \citep[][see also \citealt{DBB06}]{JmLB09}. Therefore, an increase of the galactic outflow intensity and an enhanced contribution of turbulence in the off-planar gas appear to be the dominant effects of the increased SFR on a spiral galaxy. In this context, the increase of the turbulent speed with the SFR adopted in the fiducial model can be interpreted as a reflection of the increasing importance of the off-planar gas with its more intense turbulence. 

Similarly to \citetalias{Rodrigues+19}, we assume that the rms turbulent speed is independent of the galactocentric distance $r$. Some observations of the \HI\ velocity dispersion in spiral galaxies show that it decreases with the galactocentric radius $r$ from often supersonic values in the inner parts of galaxies \citep{BV92,PeRu07,TRLmLWKBdB09,Mogotsi+16}. On the other hand, there is significant observational evidence for a very weak variation of the \ion{H}{i} velocity dispersion with $r$ in the discs of spiral galaxies \citep[\citealt{DHH90,BS91}, Section~12.2.3 of][]{Kamphius1993PhDT}. It is plausible that the turbulent speed is indeed independent of the galactocentric radius and the spatial variation of the gas velocity dispersion is due to variations in the outflow intensity. This aspect of the model requires further analysis.

The models and observational estimates of the \ion{H}{i} and H$\alpha$ velocity dispersions discussed in Section~\ref{ism_turbulence} are often interpreted as suggesting strongly supersonic turbulence, especially at high SFR. However, the turbulence in the \textit{diffuse} (warm and hot) interstellar gas is likely to be transonic (or subsonic, depending on the kinetic energy injection rate) because of the self-regulation of the supersonic turbulence \citep[Section~2.10.2 of][]{ss21}. The kinetic energy of supersonic turbulence efficiently dissipates at shock fronts to heat the gas until an equilibrium state is reached in which the turbulent speed is comparable to the sound speed. Thus, turbulent flows with sufficiently strong energy injection rates are likely to be transonic. Such a self-regulation of a turbulent system can be affected by radiative cooling which can prevent the gas heating by removing the dissipated turbulent energy via radiation (Enrique V{\'a}zquez-Semadeni, private communication). The cooling is especially efficient in dense regions (like molecular clouds which occupy a negligible fraction of the disc volume), where supersonic turbulence can be maintained, but perhaps not in the warm and hot ISM phases which dominate the radio luminosity. The self-regulation of supersonic turbulence is not included in the turbulence models of \citet{Krumholz_18} and \citet{Ginzburg_Dekel+22}, who assume that the disc is marginally stable concerning the gravitational instability at all times and consider only the energy balance of turbulence driven by supernovae and accretion flows onto and through the disc. The sample of SFGs where \citet{Varidel_Croom+20} measured the vertical velocity dispersion in the H$\alpha$ spectral line deliberately includes only nearly face-on galaxies ($0<i<60^\circ$ for the inclination angle), so not only the turbulence in the galactic discs but also outflows and turbulence of the off-planar gas are likely to contribute to the velocity dispersion obtained.

The dependence of $v$ on the SFR, equation~\eqref{eq_v0}, is rather poorly constrained by the data available, and other fits can be equally acceptable. For example, adopting  $\SFR_0=1\msun\yr^{-1}$ and $c=1/3$ in our fiducial model produces an RLF that is in reasonable agreement with observations, indicating a degeneracy between these parameters (lower values of the scaling exponent $c$ would require lower values of $\SFR_0$).  Furthermore, there are alternative prescriptions to derive the velocity dispersion from the star formation surface density \citep{Niklas_beck_1997, Chyzy+2011, Schleicher_beck+2013, Chamandy+24}. Finally, the velocity dispersion may increase faster with redshift than expected from the SFR--redshift dependence alone \citep{Wisnioski+15,Ubler+19}, so a better understanding of the interstellar turbulence and the role of the evolving multi-phase ISM structure appears to be essential for further progress in modelling the RLF, which we will address in the future.

\section{Conclusions}\label{sec_conclusion}
The magnetic field and dynamo models presented here lead to a satisfactory agreement, within uncertainties, of the predicted RLF of star-forming galaxies with observations up to $z\simeq2$. At higher redshifts, although the form of the theoretical luminosity function is similar to what is observed, the theory predicts a smaller number of galaxies of high luminosity, and the discrepancy increases with the redshift. While the observational data at higher redshifts still lack galaxies of faint and moderate luminosity, our findings enable us to identify a range of effects that need to be understood and included to improve the agreement with the data.

We have identified the strength of interstellar turbulence as one of the major factors affecting the RLF. A better understanding of the interstellar turbulence, primarily the turbulent speed (which makes only a limited and uncertain contribution to the observed spectral line widths) at high redshifts and the effects of the evolving multi-phase structure of the interstellar medium on magnetic fields and cosmic rays are required to improve our models.

Models presented here are based on a well-tested galaxy formation model; however, we rely on an implicit assumption that the overall structure of the interstellar medium does not change as the galaxies evolve. This is an oversimplification but little is known about some aspects of galactic evolution which are relevant to magnetic fields and cosmic rays (the turbulent parameters, gas disc thickness, the multi-phase ISM structure, etc.). It is then not surprising that our model becomes inapplicable beyond the redshift $z\simeq2$ at which galaxies undergo a significant structural change, the development of pronounced gas discs in particular. Among distinct structural features of young galaxies, important for their nonthermal constituents, might be widespread radio haloes associated with stronger star formation and correspondingly vigorous galactic fountain flows. Our limited knowledge of the properties of turbulent magnetic fields, especially in the evolving multi-phase ISM, is another question that requires a better answer.

We have shown that the galactic luminosity function in the total radio intensity is a sensitive indicator of the state of the interstellar gas, especially its turbulence, density, and galactic fountains and winds. Our results are consistent with the notion that the structure of star-forming galaxies, and their magnetic fields, are different at $ z\lesssim2$ and the higher redshifts. The luminosity function in polarised radio emission would provide even richer information on galactic rotation, gas stratification, and anisotropy of interstellar turbulence.
\section*{Acknowledgements}
We gratefully acknowledge the detailed, thoughtful comments of the anonymous referee. We are grateful to Dominik J.~Schwarz, Aritra Basu and Sukanta Ghosh for useful discussions. CJ is supported by the University Grant Commission, India through a BSR start-up grant (F.30-463/2019(BSR)) and the Rashtriya Uchchatar Shiksha Abhiyan (RUSA) scheme (No.CUSAT/PL(UGC).A1/2314/2023, No:T3A). CJ also acknowledges the access to the high-performance cluster at IUCAA (Pune, India) facilitated through the associateship program. CMB is supported by the UK Science and Technology Funding Council (STFC) through grants ST/T000244/1 and ST/X001075/1. 
\section*{Data Availability Statement}
The model data presented in this paper are available from the first author upon a reasonable request.  

\bibliographystyle{mnras}
\bibliography{magnetizer}

\appendix
\section{Galactic mean-field dynamo}\label{sec:MFD}
The interstellar magnetic field can be separated into mean and random components, 
\begin{equation}
  \label{B}
  \vec{B}=\mean{\vec{B}}+\vec{b},
\end{equation}
where the overbar represents ensemble or volume averaging. The mean magnetic field is amplified by the joint inductive action of the galactic differential rotation (at an angular velocity $\omega$) and helicity of interstellar turbulence (the $\alpha$-effect) in what is known as the $\alpha\omega$-dynamo \citep[][and references therein]{ss21}. The intensity of these effects relative to the turbulent magnetic diffusion, which destroys the mean magnetic field by tangling it, is quantified by the dimensionless dynamo number,
\begin{equation}\label{DSA}
  D=\frac{\alpha S h\disc^3}{\beta^2}\,,
\end{equation}
where $\alpha$ is the magnitude of the $\alpha$ effect, $\beta$ is the turbulent magnetic diffusivity and $S=r\,\dd\omega/\dd r$.
The induction effects that amplify the magnetic field can overcome the destructive action of the turbulent diffusion if the magnitude of the dynamo number exceeds a certain critical value $|D\cra|$, $|D|>|D\cra|$, and we note that both $D<0$ and $D\cra<0$ because $S<0$ in most parts of galactic discs. In a thin, rotating gas layer, $D\cra\approx-10$. Using the estimates $\alpha\simeq l_0^2\omega/h\disc$ and $\beta\simeq\tfrac13 l_0 v$, where $l_0$ and $v$ are the turbulent scale and speed, the dynamo number can be expressed in terms of directly observable galactic parameters,
\begin{equation}\label{DOH}
D\simeq 9\frac{\omega Sh\disc^2}{v^2}\,.
\end{equation}
The magnetic field thus generated in a thin disc has the quadrupolar parity with respect to the galactic mid-plane $Z=0$: $\mean{B}_r(-Z)=\mean{B}_r(Z)$, $\mean{B}_\phi(-Z)=\mean{B}_\phi(Z)$ and $\mean{B}_z(-Z)=-\mean{B}_z(Z)$ in cylindrical coordinates $(r,\phi,Z)$.

As the magnetic field strength increases to become comparable to $B\eq$, the Lorentz force becomes comparable to the Coriolis and other forces, and the field amplification slows down until the dynamo reaches a saturated, steady state where the magnetic field strength is estimated as \citep[][ and sections 12.3 and 13.7.3 of \citealt{ss21}]{Chamandy+14b}
\begin{equation}\label{Bss}
\mean{B}^2\simeq \uppi^2 B\eq^2\left(\frac{l}{h\disc}\right)^2\left(\frac{D}{D\cra}-1\right)R_\kappa\,.
\end{equation}
where $R_\kappa$ is a parameter of order unity, and we have set the outflow speed to zero. \magnet\ includes the nonlinearity associated with the magnetic helicity balance and solves for the equivalent strength of the magnetic field as a function of the galactocentric radius $r$ using the so-called no-$Z$ approximation applicable to a thin disc, $h\disc/r\lesssim0.1$. The magnetic field components in this solution can be thought of as the mid-plane values if the magnetic field distribution in $Z$ is exponential (and the average values otherwise).

\section{Synchrotron luminosity }
\label{appen_syn_lum_1D}
\begin{figure}
\centering
\includegraphics[width=0.48\textwidth]{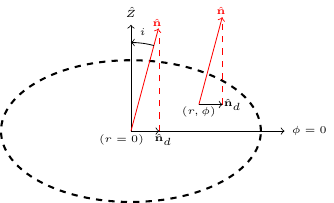}
\caption[ LoS]{The schematic diagram of a galaxy disc plane showing the line of 
sight unit vectors $\uvec{n}$ passing through $r=0$ and an arbitrary location $(r, \phi)$ as red arrows. The vector $\uvec{n}\disc$ is the projection 
of $\uvec{n}$ onto the disc plane.  
}
\label{fig_galaxy_los}
\end{figure}
The synchrotron luminosity given in equations~\eqref{eq_IB} and \eqref{eq_syn_Lum} is proportional to the integral $\int_{V} K_E B_\perp^2 \, \dd ^3\vec{r}' \,,$ where $K_E\propto B^2$ under the assumption of the energy equipartition between cosmic rays and the magnetic field. 
\magnet\ solves the mean-field dynamo equations for the cylindrical components of the axisymmetric mean magnetic field $\vec{\mean{B}}=(\mean{B}_r, \mean{B}_\phi, \mean{B}_z)$ for each galaxy in the sample as a function of the distance to the galactic centre $r$ in the galactic disc plane (corresponding to $Z=0$), and we augment the solution assuming its exponential distribution across the disc (in $|Z|$) with the scale height $h_B$ (Section~\ref{sss_B_Z_dist}). The random magnetic field is assumed to have the same scale height, with the strength obtained from the turbulent energy density (Section~\ref{sec:random}). 

The synchrotron luminosity depends on the inclination angle $i$ ($0 \leq i \leq \pi$) between the observer's line of sight (LoS) and the $Z$-axis (perpendicular to the galactic disc). The  unit vector along any LoS ($\uvec{n}$) is resolved into  $\vec{n}\disc = \uvec{n}\disc \sin{i}$ 
along the galactic disc plane and $\vec{n}_Z = \uvec{n}_Z \cos{i} $ along the $Z$-axis (see Fig.~\ref{fig_galaxy_los}). The $\phi=0$ direction is  determined by the direction of $\uvec{n}\disc$ as it passes through $r=0$. Hence, the $r$ and $\phi$ components of $\uvec{n}_d$ for an arbitrary 
LoS passing through  a location $(r,\phi,0)$ in the disc plane are $\cos{\phi}$ and $-\sin{\phi}$ and correspondingly, the components of the unit 
vector $\uvec{n}$ passing through  $(r,\phi,0)$ are 
\begin{equation}\label{nhat}
   \uvec{n} \equiv (n_r, n_\phi, n_Z) = (\sin i \cos \phi, -\sin i \sin \phi,   \cos i)\,,  
\end{equation}
which is used to derive the magnetic field component in the sky plane $\vec{\mean{B}}_\perp = \vec{\mean{B}}-\uvec{n}(\vec{\mean{B}}\cdot\uvec{n})$. 

Assuming that the turbulent magnetic field $\vec{b}$ is isotropic, $\mean{b_\perp^2}=\tfrac23 b\rms^2$, and $b\rms$ is estimated in Section~\ref{sec:random}. Since we have adopted $K_E\propto\mean{B}^2=\mean{B}^2+b\rms^2$, the synchrotron luminosity of a galaxy is proportional to the following integral over the volume $V$ of the magnetised region:
\begin{equation}\label{Jint3}
J=\int_{V} (\mean{B}^2+b\rms^2) (\mean{B}_\perp^2+\tfrac23  b\rms^2) \, \dd ^3\vec{r}' \,.
\end{equation}
When both parts of the magnetic field have an exponential distribution in $Z$ with a scale height $h_B$ and integration over $Z$ is extended over $|Z|<\infty$, we have
\begin{equation}\label{Jint}
J= \dfrac{1}{2} \int_0^{r\disc}r'\dd r'\int_0^{2\pi}\dd\phi\, h_B(r') (\mean{B}^2+b\rms^2) (\mean{B}_\perp^2+\tfrac23 b\rms^2) \,,
\end{equation}
where we adopt $h_B=h\disc$; $h\disc$ and $r\disc$ are defined in Sections~\ref{DGQ} and \ref{sss_B_Z_dist} and $\mean{B}_\perp$ depends on $\phi$ via the unit LoS vector \eqref{nhat}. We note that the scale height of the magnetic field can be larger than that of the gas. For example, $h_B=2h\disc$ for $B\propto B\eq\propto\rho^{1/2}$. Such an increase in $h_B$ would have led to an increase in the RLF by a factor of two, so the values of the RLF presented are conservative.

\section{Testing the synchrotron emission model: M51 galaxy}
To verify the evaluation of the synchrotron luminosity, we apply the procedure described above to the nearby spiral galaxy M51. The radio emission of this galaxy is thoroughly explored and interpreted in terms of the galactic magnetic field by \citet{FBS11}. To derive the magnetic field strength, these authors use the assumption of the local energy equipartition between cosmic rays and magnetic fields, similar to equation~\eqref{eq_crBeq}. 
The total magnetic field strength derived from the radio intensity at the wavelength $\lambda=6\cm$  ($\nu=5\GHz$ in terms of the frequency) assuming the path length of $S =1\kpc$ through the synchrotron-emitting region is $30\mkG$ in the central part of the galaxy, 20--$25\mkG$ in the spiral arms and 15--$20\mkG$ between the arms in the main part within $5\kpc$ of the centre \citep[section~4.1 and fig.~8 of][]{FBS11}.
The mean magnetic field strength derived from the Faraday rotation does not exceed $3\mkG$ \mbox{\citep[section~6.2 of][]{FBS11}}.
The intensity of the radio emission at $\lambda=6\cm$ for distances $1.6<r<4.8\kpc$ from the centre, given in Table~2 of \citet{FBS11}, ranges from $0.5\mJyb$ between the arms to $1.1\mJyb$ within them, with a beam of $W=15\,\text{arcsec}$ in diameter. It includes the thermal emission, and its fraction at $\lambda=6\cm$ is estimated as 25 per cent \citep[section~4.1 of][]{FBS11}.

Our goal is to test our choice of various parameters involved in the calculation of the synchrotron intensity rather than to achieve a precise agreement with the observations of M51. Therefore we only include isotropic and homogeneous magnetic field (with no $Z$ dependence) with a scale-height of $0.5\kpc$ (half the path length of the synchrotron-emitting region) that is independent of the distance from the galactic centre. The rest of the parameters are the same as in the fiducial model presented in this paper. The synchrotron intensity is derived using equations~\eqref{eq_IB} and \eqref{eq_syn_Lum} with $b_\perp^2 =\tfrac{2}{3} b^2$ neglecting both the anisotropy of the random field and the mean magnetic field since $b^4\gg \mean{B}^4$ in M51; this makes our estimate of the radio intensity quite conservative.

The total synchrotron intensity  for the  solid angle $\pi(W/2)^2$ of a flat beam is then
\begin{equation} \label{I_toy}
    I(\nu) = \pi \left( \frac{W}{2} \right)^2   \frac{L(\nu)}{4 \pi }  \epsilon_\text{th} \,, 
\end{equation}
where the factor $\epsilon_\text{th}=1.25$ accounts for the contribution of the thermal radio emission. The total synchrotron intensity from \eqref{I_toy} at $\nu = 5\GHz$ in the main part of the disc of M51 within $r= 5\kpc$ for a field strength of $25\mkG$, and a flat beam of $W=15\,\text{arcsec}$ in diameter is $0.7\mJyb$, which is in good agreement with \cite{FBS11}, who measures the same to be $0.8\mJyb$.
If we instead use the magnetic field strength of $20$ $\mu$G in our model, the resulting synchrotron intensity is $0.3\mJyb$.       

\label{lastpage}
\bsp	
\end{document}